\documentclass[12pt]{iopart}

\usepackage{iopams}
\usepackage{graphicx}  
\begin{document}

\title[Negative index and gold nano-checkerboards]{Plasmonic interaction of light with negative index and gold nano-checkerboards}

\author{Sangeeta Chakrabarti}
\address{Department of Physics, Indian Institute of Technology,
Kanpur 208016, India \\
Institut Fresnel, UMR CNRS 6133, Aix-Marseille Universit\'e,\\
Campus universitaire de Saint-J\'er\^ome, 13397 Marseille, France}
\author{S. Anantha Ramakrishna and Neeraj Shukla}
\address{Department of Physics, Indian Institute of Technology,
Kanpur 208016, India}
\author{Fanny Guenneau, Muamer Kadic, Sebastien Guenneau and Stefan Enoch}
\address{Institut Fresnel, UMR CNRS 6133, Aix-Marseille Universit\'e,\\
Campus universitaire de Saint-J\'er\^ome, 13397 Marseille, France}
\ead{sar@iitk.ac.in}
\begin{abstract}
Negative refractive index materials (NRIM) make possible unique
effects such as a convergent flat lens due to the reversed
Snell-Descartes laws of refraction. NRIM are also known to
be able to support a host of surface plasmon states for both
polarizations of light which are responsible for the sub-wavelength
image resolution achieved by a slab of NRIM.
A generalized lens theorem provides us with a class of
spatially varying slab lenses satisfying the prerequisite symmetries
to fold the optical space onto itself. This theorem can be derived using
powerful tools of transformational optics. A paradigm of such
complementary media are checkerboards consisting of alternating
cells of positive and negative refractive index that represent a
very singular situation in which the local density of modes at
the corners are enormously enhanced. We have considered several
theoretical and numerical aspects of such structured films
including finite slabs of multi-scale checkerboards of NRIM
satisfying the generalized lens theorem which are host of strongly
enhanced electromagnetic field. Such checkerboards can be mapped
using transformational optics onto three-dimensional corner lenses
consisting of semi-infinite heterogeneous anisotropic regions of space
satisfying the generalized lens theorem.
It is also possible to design three-dimensional checkerboards of complementary
media, the only restriction being that corresponding unfolded structures in the
plane are constrained by the four color theorem.
Some of these meta-surfaces in the plane display
thin bridges of complementary media, and this further enhances their
plasmonic response. Since plasmonic metals mimic the behaviour
of NRIM at small length scales, opaque gold
films structured at sub-micron scales in a checkerboard fashion were
fabricated using focussed-ion-beam technologies and their scattering spectra measured.
Sub-wavelength square holes in a thick gold film placed in checkerboard fashion
show a broadband extra-ordinary transmission of light.
These structures are seen to have enhanced interaction of light
at the edges and corners. There is a strong correspondence between
the theoretical predictions and the experimental measurements.
\end{abstract}

\maketitle

In 1967, Veselago proposed a thought experiment in which materials
with simultaneously negative permittivity
($\varepsilon$) and magnetic permeability ($\mu$) were shown to have
a negative refractive index~\cite{veselago}. A ray analysis allowed
him to conclude that a slab of such a negative refractive index
material (NRIM) can act as a flat lens that imaged a source on one
side to a point on the other. But this result remained an academic
curiosity for almost thirty years, until Pendry and
co-workers~\cite{pendry96,pendry_IEEE} proposed designs of structured
materials which would have negative effective $\varepsilon$ and
$\mu$. Further, Pendry also showed that the flat lens proposed by Veselago
was very unusual in that the image resolution produced by this lens in
principle, did not have any limitation~\cite{pendry_prl00}.
These so-called meta-materials are indeed structured at sub-wavelength length scales
(typically $\lambda/10~\mathrm{to}~\lambda/6$),
hence it is possible to regard them as almost homogeneous. The first
experimental realizations were chiefly achieved at GHz frequencies
~\cite{smith00,parazzoli}, but meta-materials in the near infrared
and optical frequencies have been proposed and demonstrated.
More recently, new solutions based on geometric transformations
to the material parameters and Maxwell's equations in curvilinear coordinate systems, reported
by Greenleaf {\it et al.}~\cite{greenleaf}, Pendry {\it et al.}
~\cite{pendry} and Leonhardt~\cite{leonhardt06}, have paved the way
towards a markedly enhanced control of electromagnetic waves around
arbitrarily sized and shaped solids, leading to electromagnetic invisibility, even in the extreme near field
limit~\cite{zolla}. The experimental validation of these theoretical
considerations has been given by Schurig {\it et al.}~\cite{pendryexp}, who
used a cylindrical cloak consisting of concentric arrays of split
ring resonators.


However, the touchstone of research in metamaterials remains the
quest for the perfect lens: in a seminal paper, Pendry demonstrated
that the Veselago slab lens not only involves the propagative waves
but also the evanescent near-field components of a source in the
image formation~\cite{pendry_prl00}. Such a superlensing effect has
been demonstrated at optical frequencies through a silver slab film
in~\cite{zhang_science05} (resolution of $\lambda/5$). It was
shown by Pendry and Ramakrishna\cite{pendry_jpc03} that the
superlensing effect with a slab of negative refractive index medium
can be generalized  to materials that are anisotropic and spatially inhomogeneous.
The only condition is that the system has to respect a mirror
anti-symmetry about a plane normal to the imaging axis. Using a
geometric technique it was shown~\cite{pendry_jpc03}, as a
consequence of this theorem, that two rectangular (semi-infinite)
intersecting wedges of NRIM acts as an imaging system whereby a
source gets imaged onto itself. This system, originally studied by
Notomi~\cite{notomi} using a ray picture, was thus shown to involve
the evanescent modes also and act as a unique resonator. Guenneau
{\it et al.}~\cite{guenneau_njp05} subsequently generalized this
imaging effect to a rectangular checkerboard lattice where
alternating cells have positive ($\varepsilon = \mu = +1$) and
negative ($\varepsilon = \mu = -1$) refractive index. It was shown
that a source placed in one cell would reproduce itself in every
other cell of the infinite lattice. The properties of corners and
checkerboards in the presence of dissipation have also been
studied using geometric transforms ~\cite{guenneau_ol05,sangeeta}.
These transformational optics tools are reminiscent of the work by
Leonhardt and Philbin on multi-valued maps for lensing effects via
negative refraction \cite{philbin} which were further investigated
in \cite{crp2009}.
Monzon {\it et
al.}~\cite{cesar} recently derived an analytical solution for a
finite sized NRIM wedge in the presence of a source. He {\it et al.}
~\cite{he_njp} studied some modes of a resonator with NRIM wedges and
constructed an open cavity using triangular wedges of a PC that
shows the negative refraction effect (Also see \cite{pra2007}).

In a parallel development in 1998, Ebbesen et al. established that resonant excitation of
surface plasmons enhance electric fields at a surface that force
light through its tiny holes, giving very high transmission
coefficients in the sub-wavelength regime~\cite{ebbesen}. Pendry,
Martin-Moreno and Garcia-Vidal further showed in 2004 that one can
manipulate surface plasmon ad libitum via homogenization of
structured surfaces~\cite{science2004}. In the same vein, pioneering
approaches to invisibility relying upon plasmonic metamaterials have
already led to fascinating results
~\cite{milton2,engheta,javier,baumeier}. These include plasmonic
shells with a suitable out-of-phase polarizability in order to
compensate the scattering from the knowledge of the electromagnetic
parameters of the object to hide, and external cloaking, whereby a
plasmonic resonance cancels the external field at the location of a
set of electric dipoles. Recently, Baumeier {\it et al.} have demonstrated
theoretically and experimentally that it is possible to reduce
significantly the scattering of an object by a surface plasmon
polariton, when it is surrounded by two concentric rings of point
scatterers~\cite{baumeier}.

It is now well-known that metals that have negative permittivity can mimic
the electromagnetic properties of NRIM in the extreme near-field, i.e.,
when all the length scales in the system are small compared to the
wavelength of light~\cite{pendry_prl00}. In fact, the imaging by a
superlens at optical frequencies was demonstrated through a silver
slab film in~\cite{zhang_science05} with a  resolution of about
$\lambda/5$.This naturally prompts the question as to whether
plasmonic gold and silver nano-films
structured as checkerboards can enable control of light and
enhance light transmission through the nano-hole apertures while
confining light within the transverse plane.

Here we discuss the interaction of light with finite
checkerboard structures of NRIM and show that light strongly interacts,
namely the corners and the edges. New varieties of checkerboard-like
patterns that we call {\it origami lenses} are presented. The difficulties
of numerically calculating the properties of such checkerboard systems
are explained. We have utilized focussed ion beam technologies for fabricating submicron
sized metallic structures. A recently developed ion-beam irradiation
assisted  adhesion technique~\cite{neeraj_nimb2009} was used for making
structures where only small patterned regions are retained.
We demonstrate the power of this technique here in making complex structures
such as checkerboards on gold films. Spectral measurements of the fabricated
checkerboard structures are presented. From a comparison of the bright and
dark field images and spectra of the structures, it is concluded that most
of the scattering originates in the edges and corners of the structure.
Broadband near-infra-red transmission through sub-wavelength holes in a gold film
 arranged in checkerboard fashion is demonstrated.

\section{Singular electromagnetics of checkerboard structures}

\subsection{Generalized Lens Theorem: Complementary Media}
\label{genl}

\begin{figure}[tb]
\begin{center}
\scalebox{0.7}{\includegraphics[width=10cm,angle=-0]{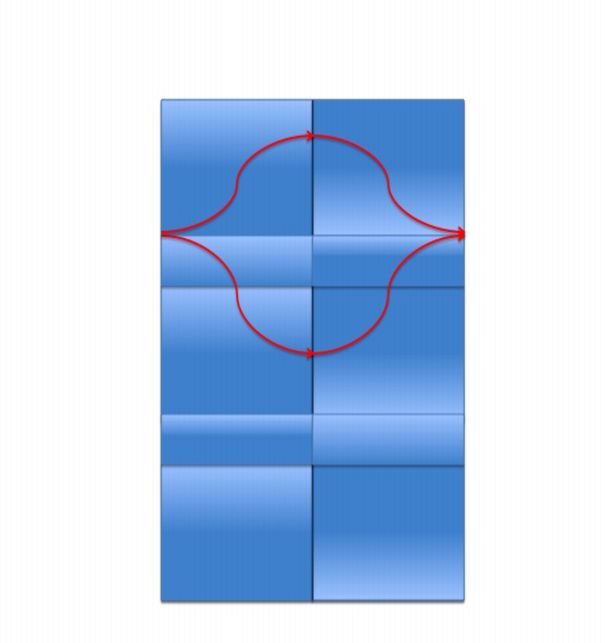}}
\scalebox{0.7}{\includegraphics[width=10cm,angle=-0]{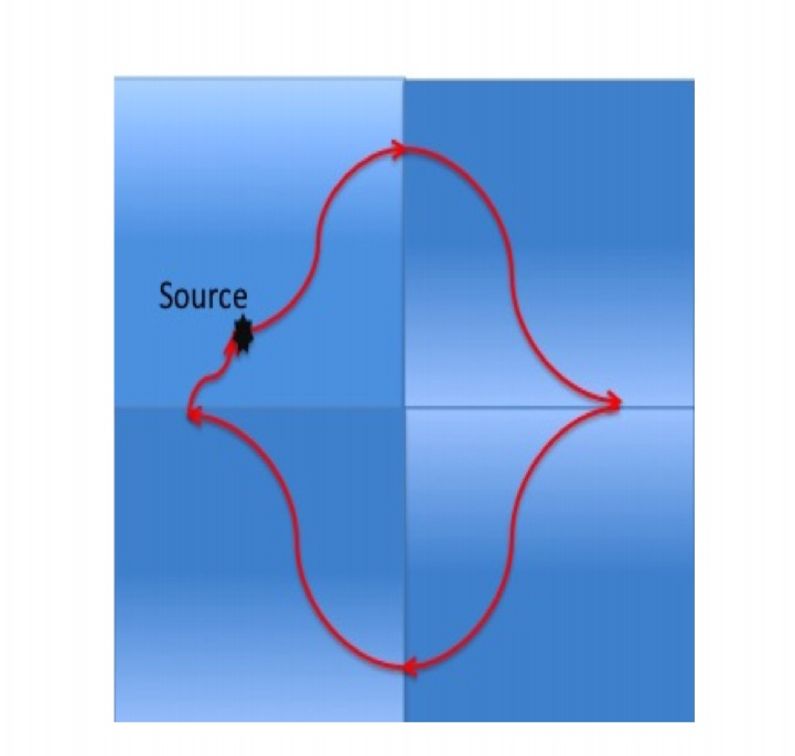}}
\caption{Left panel: Principle of the generalized lens theorem: A slab of spatially
varying negative refractive index optically cancels the presence of a slab of
spatially varying positive refractive index of equal thickness, provided that
condition (\ref{trick}) is met. The ray trajectories
emitted by a source on the leftmost side of the generalized lens,
whatever complex, always emerge on the rightmost side of the lens, in a
symmetric fashion.
Right panel: Principle of the perfect corner reflector: any
ray emitted by a source follows closed trajectories (the optical
space is folded back onto the source).
} \label{gen_lens}
\end{center}
\end{figure}

\begin{figure}[tb]
\begin{center}
\scalebox{1.0}{\includegraphics[width=10cm,angle=-0]{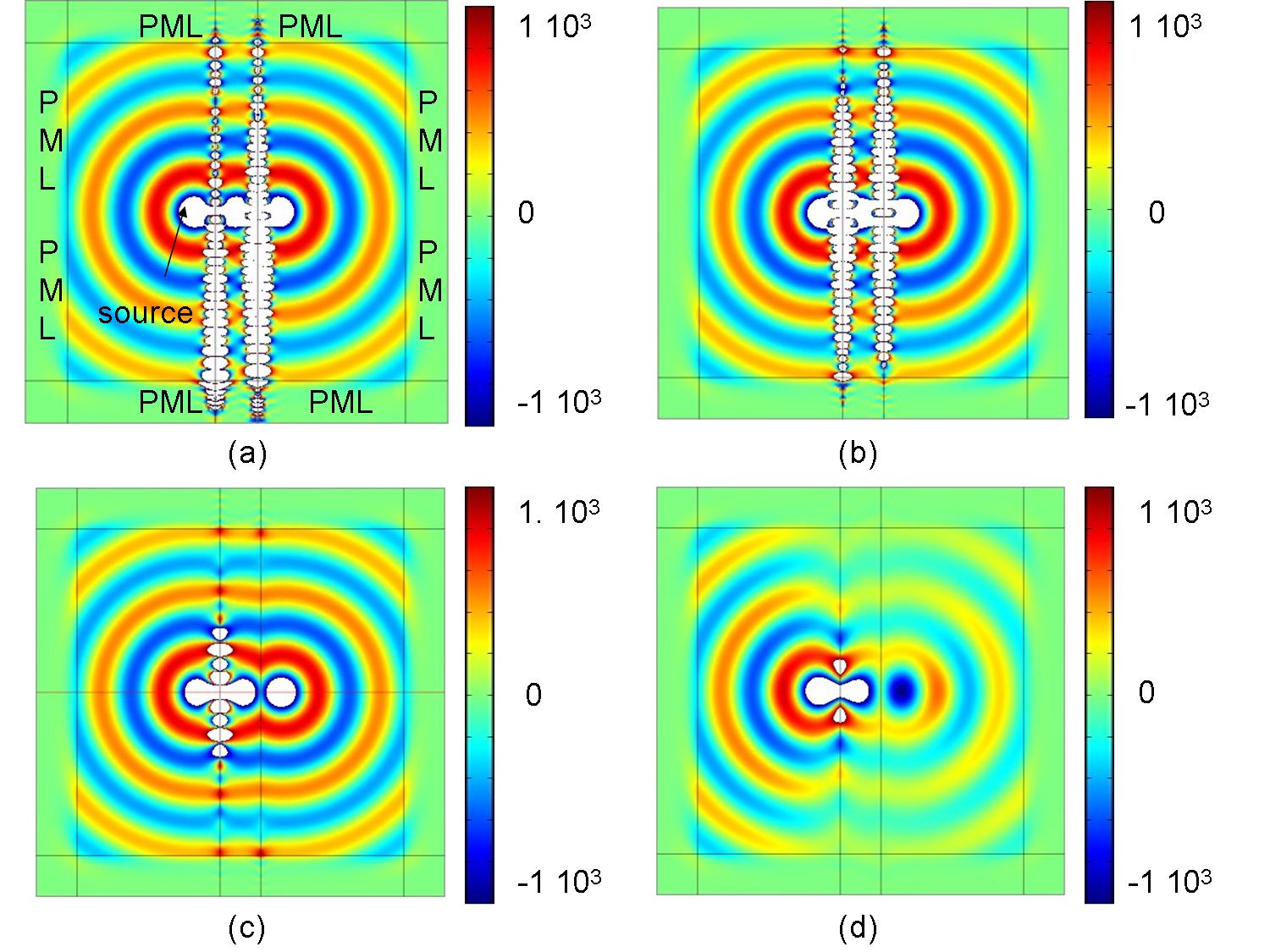}}
\caption{A point source located a distance $d/2$ away from the left
interface of a perfect lens of thickness $d$ displays an image a
distance $d/2$ away from the right interface of the lens. (a)
$\varepsilon=\mu=-1$; (b) $\varepsilon=-1+\i 10^{-6}$, $\mu=1$; (c)
$\varepsilon=-1+\i 10^{-2}$, $\mu=1$; (d) $\varepsilon=-1+\i
10^{-1}$, $\mu=1$. The results shown are for p-polarized light and $d \simeq \lambda/10$.
Perfectly Matched Layers (PMLs) consist
of either positive or negative anisotropic media depending upon
whether they model regions of infinite extent filled with positive
or negative refractive index material. } \label{s10}
\end{center}
\end{figure}

\begin{figure}[tb]
\begin{center}
\scalebox{1.0}{\includegraphics[width=10cm,angle=-0]{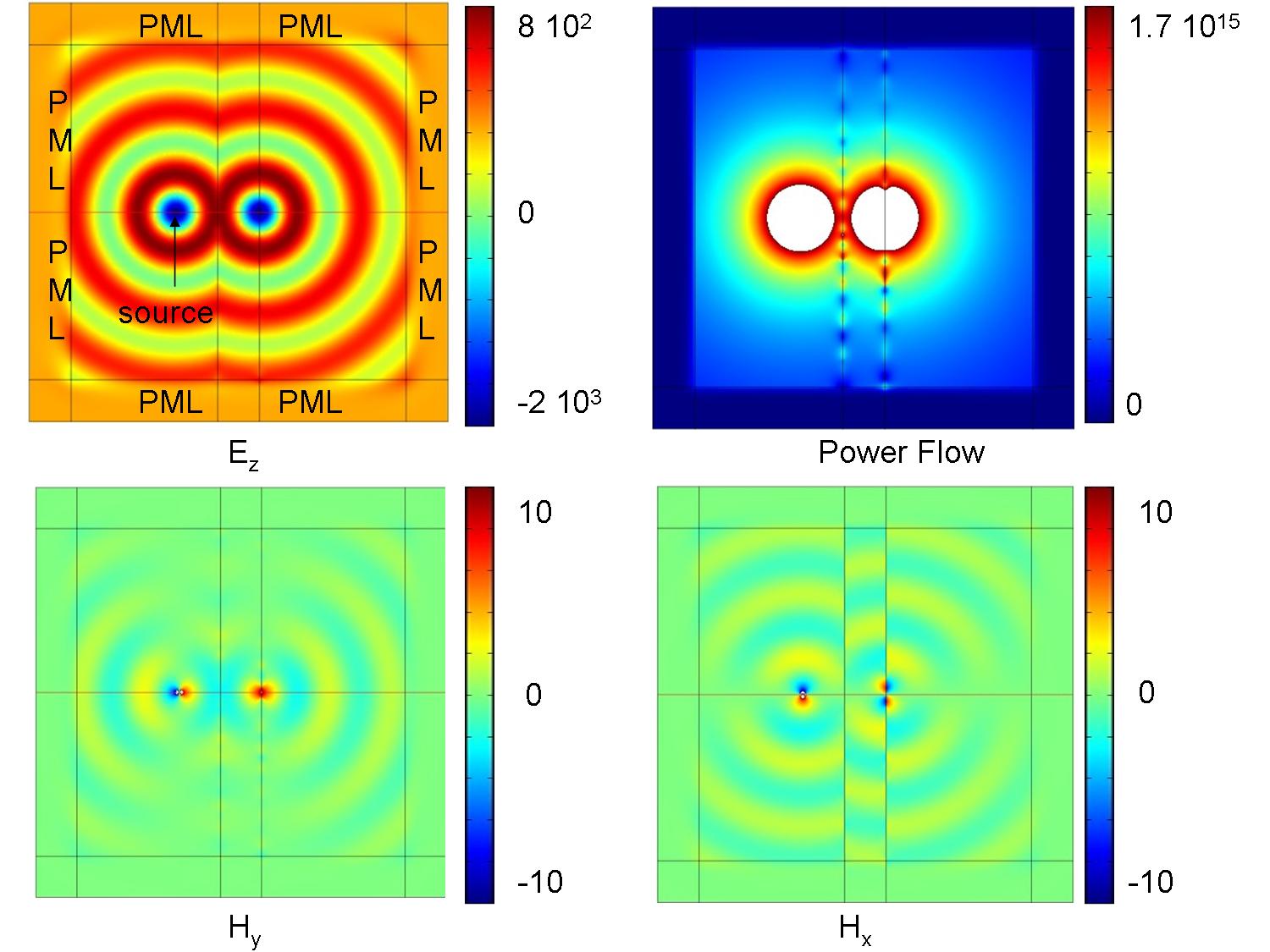}}
\caption{A point source located at a distance $d$ away from the left
interface of a perfect lens of thickness $d$ displays an image at the right interface.
(a) Longitudinal component $E_z$; (b) Power flow (c)
and (d) $H_y$ and $H_x$, respectively. The figure shows the results obtained for p-polarized light and $d \simeq \lambda/10$.} \label{s11}
\end{center}
\end{figure}

Some of the properties of checkerboard structures can be deduced by
resort to the so-called generalized lens theorem, as illustrated in
Fig.~\ref{gen_lens}. The original `perfect lens' presupposed a slab of material
with $\varepsilon$ =-1 and $\mu$ = -1, as shown on Fig.~\ref{s10}(a), whereby a source
is mirror imaged on the other side of the slab lens. In Fig.~\ref{s11}, we note that such a
lens is simply shifting the location of the source plane by a distance $d$, where $d$ is the thickness of
the lens: a source located on the left interface of the lens produces an image of the right interface.
This property is in fact very general and is the cornerstone of any imaging system via
complementary media: the image plane is mapped onto the source plane.

However, focussing will occur under more general conditions. Any
system for which
\begin{equation}
\begin{array}{cccc}
&\varepsilon_1 = + \varepsilon(x,y), &\mu_1 = +\mu(x,y), &-d < z <
0, \nonumber \\
&\varepsilon_2 = - \varepsilon(x,y), &\mu_2 = -\mu(x,y), &0 < z < d
\end{array}
\label{trick}
\end{equation}
will show identical focussing. Focussing will always occur
irrespective of the medium in which the lens is embedded. This is
true for any medium which is mirror antisymmetric about a plane.
Thus, in general, a negatively refracting medium is complementary to
an equal thickness of vacuum and optically 'cancels' its
presence.The compensating action extends to both the evanescent as
well as the propagating modes. Due to this,there is perfect
transmission and the phase change of the transmitted wave is zero.
In fact, the most general conditions include anisotropic materials
as well~\cite{pendry_jpc03}.

\begin{figure}[tb]
\begin{center}
\scalebox{1.0}{\includegraphics[width=10cm,angle=-0]{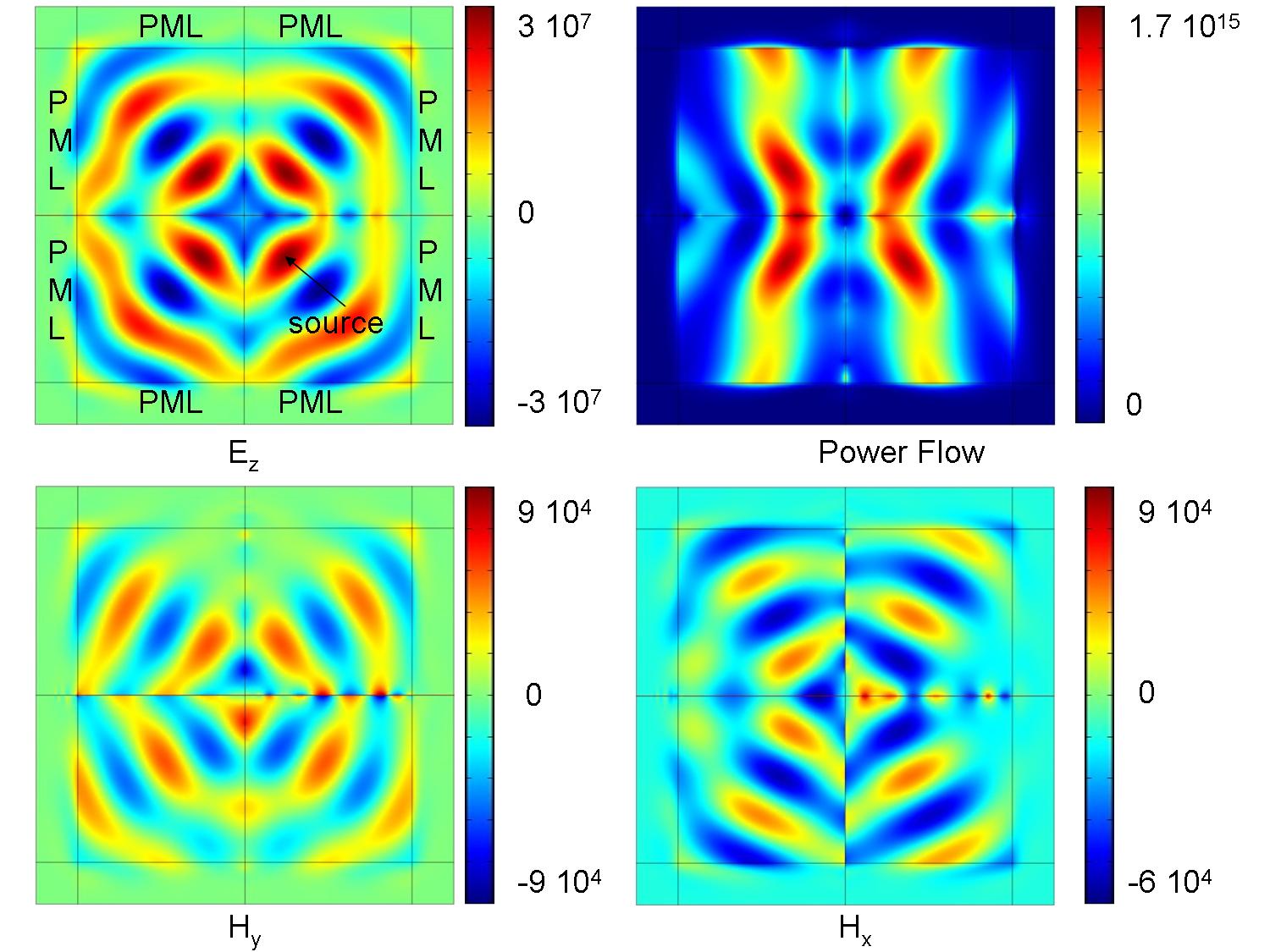}}
\caption{A point source located inside a perfect corner reflector
consisting of four infinite regions alternating positive and
negatively refracting isotropic homogeneous media
($\varepsilon=\mu=\pm 1$) displays three perfect images (one in each
corner). In such a system, light goes around in closed
trajectories and modes are infinitely degenerate, leading to an infinite
Local Density Of States (LDOS). The large magnitude of
the longitudinal electric field compared to the transverse
magnetic field is noted.
 The working wavelength is 0.3 m. Perfectly Matched Layers (PMLs) alternate positive
and negative anisotropic media depending upon
whether they model regions of infinite extent filled with positive
or negative refractive index material. } \label{s7}
\end{center}
\end{figure}

\begin{figure}[tb]
\begin{center}
\scalebox{1.0}{\includegraphics[width=10cm,angle=-0]{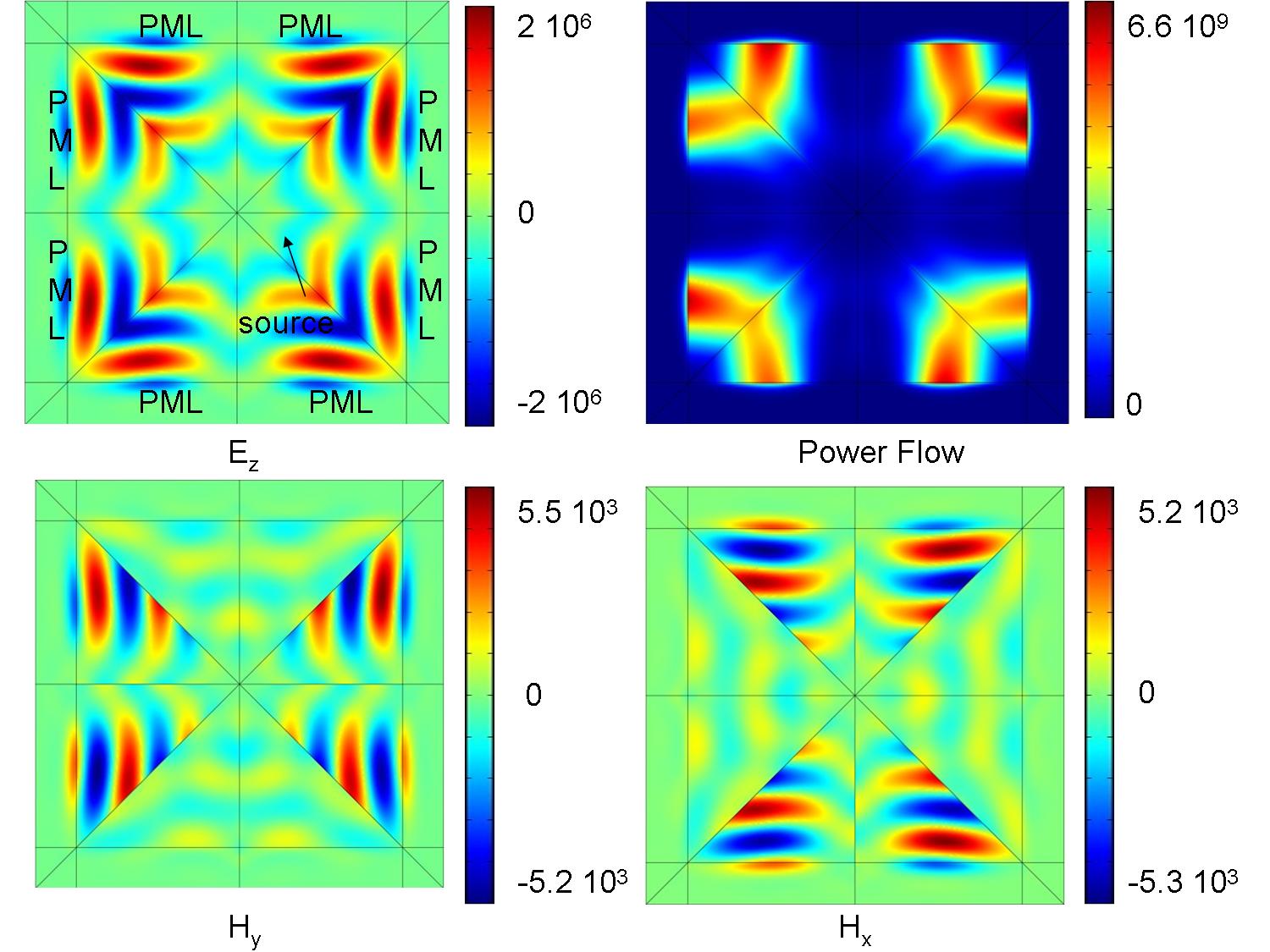}}
\caption{A point source located inside a perfect corner reflector
consisting of eight infinite regions alternating positive and
negatively refracting isotropic homogeneous media
($\varepsilon=\mu=\pm 1$) displays seven perfect images (one in each
corner). In such a system, light goes around in closed
trajectories and modes are infinitely degenerate, leading to an infinite
Local Density Of States (LDOS). The large magnitude of
the longitudinal electric field compared to the transverse
magnetic field is noted. The working wavelength is 0.3 m. It is interesting to note that the power flow
is apparently outwardly directed.} \label{s9}
\end{center}
\end{figure}

\begin{figure}[tb]
\begin{center}
\scalebox{1.0}{\includegraphics[width=10cm,angle=-0]{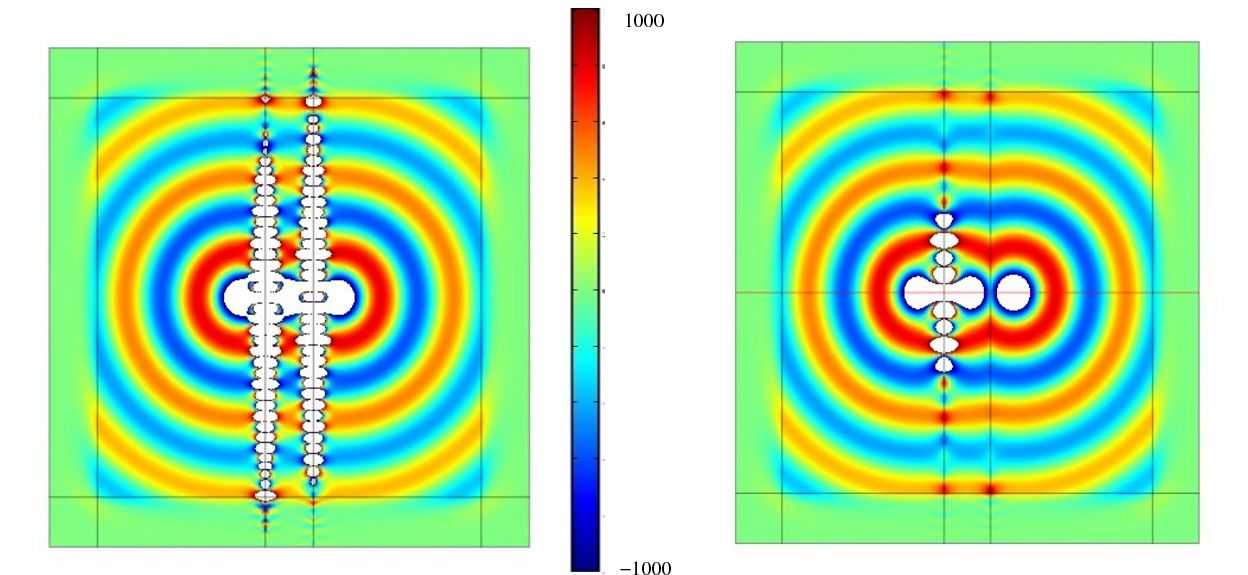}}\\
\scalebox{0.7}{\includegraphics[width=10cm,angle=-0]{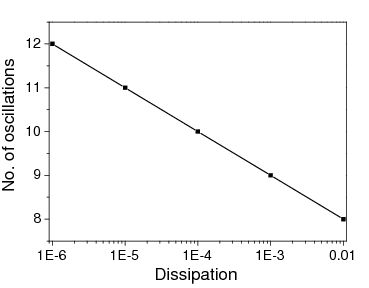}}
\caption{The dependence of the plasmon oscillations on the dissipation in the medium. For the slab lens (top), the number of
spatial oscillations is evidently lowered as the dissipation is increased from $10^{-6}$ to $10^{-3}$.
In the case of the perfect corner reflector (bottom) shown in Fig.~\ref{s9}, the number of oscillations varies inversely
with the logarithm of the dissipation in the medium.} \label{logplots}
\end{center}
\end{figure}

A corner made of a NRIM shares the perfect property of other
negatively refracting lenses. This has been shown in
~\cite{pendry_jpc03} using the technique of coordinate
transformation. A pair of negatively refracting corners with
$\epsilon = -1$  and $\mu = -1$ is capable of bending light in a
loop and forming a series of images such that the light circulates
within the loop forever (see Fig.~\ref{s7}). In the electrostatic
limit, all the surface plasmon modes in this system are degenerate
and the density of states diverges. If we can further divide each
region into two regions sharing an interface along the main
diagonals $x=\pm y$ so that we now have eight infinite regions, seven
perfect images (one in each corner) are formed(see Fig.~\ref{s9}). As in the case of
Fig.~\ref{s7}, the light circulates within the loop forever. In Fig.~\ref{logplots}, we have plotted
the number of plasmon oscillations in the medium as a function of the logarithm of the dissipation in the medium,
as expected~\cite{guenneau_ol05}. The number of oscillations depends inversely upon the logarithm of the dissipation,
which is known to affect the resolution of the image formed~\cite{merlin_apl,gomez_santos,smith_apl} as
per: $R=-2\pi d/\ln(\sigma/2)$, where $d$ is the thickness of the slab lens and $\sigma$ is the dissipation.
Clearly, the lower $\sigma$, the larger $R$, which is also
in accordance with the fact the lower $\sigma$, the larger the number of spatial oscillations of surface plasmons
at the slab interface, see also Fig.~\ref{logplots}.

A checkerboard is essentially a collection of periodically placed corners
between positive and negative refractive media. The corners and edges
are expected dominate all the optical properties of the checkerboards.
In infinite non-absorptive checkerboard structures, it has been
shown in Ref.~\cite{sangeeta} that a source placed placed in one cell of
the checkerboard produces an image in every other cell. These
checkerboard structures retain their image transfer properties
irrespective of whether they consist of homogeneous isotropic media
or inhomogeneous anisotropic media,as long as they exhibit mirror
antisymmetry and adjacent cells are complementary to each other.

All modes are degenerate at a given frequency and the density of
modes is infinite. These systems are extremely singular and contain a
very large number of corners between positive and negative cells
where the density of surface plasmon states diverges. In the absence
of dissipation, the infinite lattices of such checkerboard systems are
indeed very singular and we can only use the idea of complementary
media to deduce anything about the system. Dissipation affects
sub-wavelength imaging badly, and the divergence in the local
density of states can only worsen the situation.It is due to
this fact that the effect of dissipation on sub-wavelength
resolution becomes an important issue.

\subsection{Contradictions between the ray and the wave pictures}

\begin{figure}[h]
\begin{center}
\scalebox{1.0}{\includegraphics[width=10cm,angle=0]{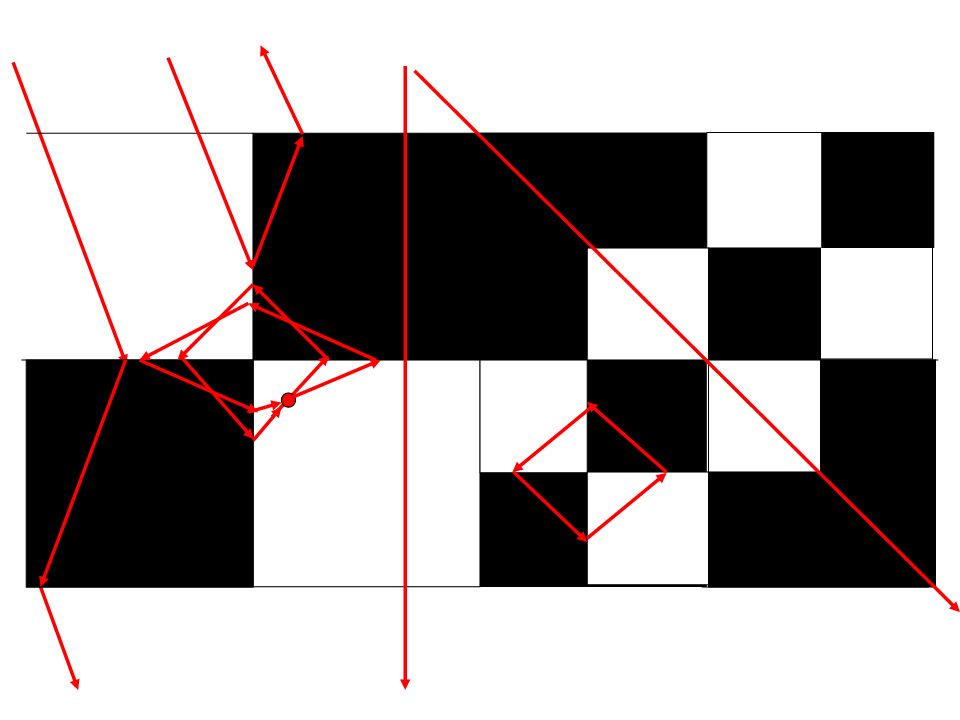}}
\caption{Checkerboard lens consisting of square and L-shaped
inclusions. Black regions have a negative refractive index, and white
regions a positive refractive index. Some incoming rays (from left)
are transmitted while others are reflected or trapped around corners
(attractors) inside the lens. This contradicts the generalized lens
theorem which predicts a full transmission. Interestingly, most rays
emanating from a point source inside the lens describe closed
trajectories (giving rise to one perfect image and two ghost
images).} \label{chess5}
\end{center}
\end{figure}

\begin{figure}[h]
\begin{center}
\hspace{0cm}\mbox{}
\scalebox{1.0}{\includegraphics[width=10cm,angle=0]{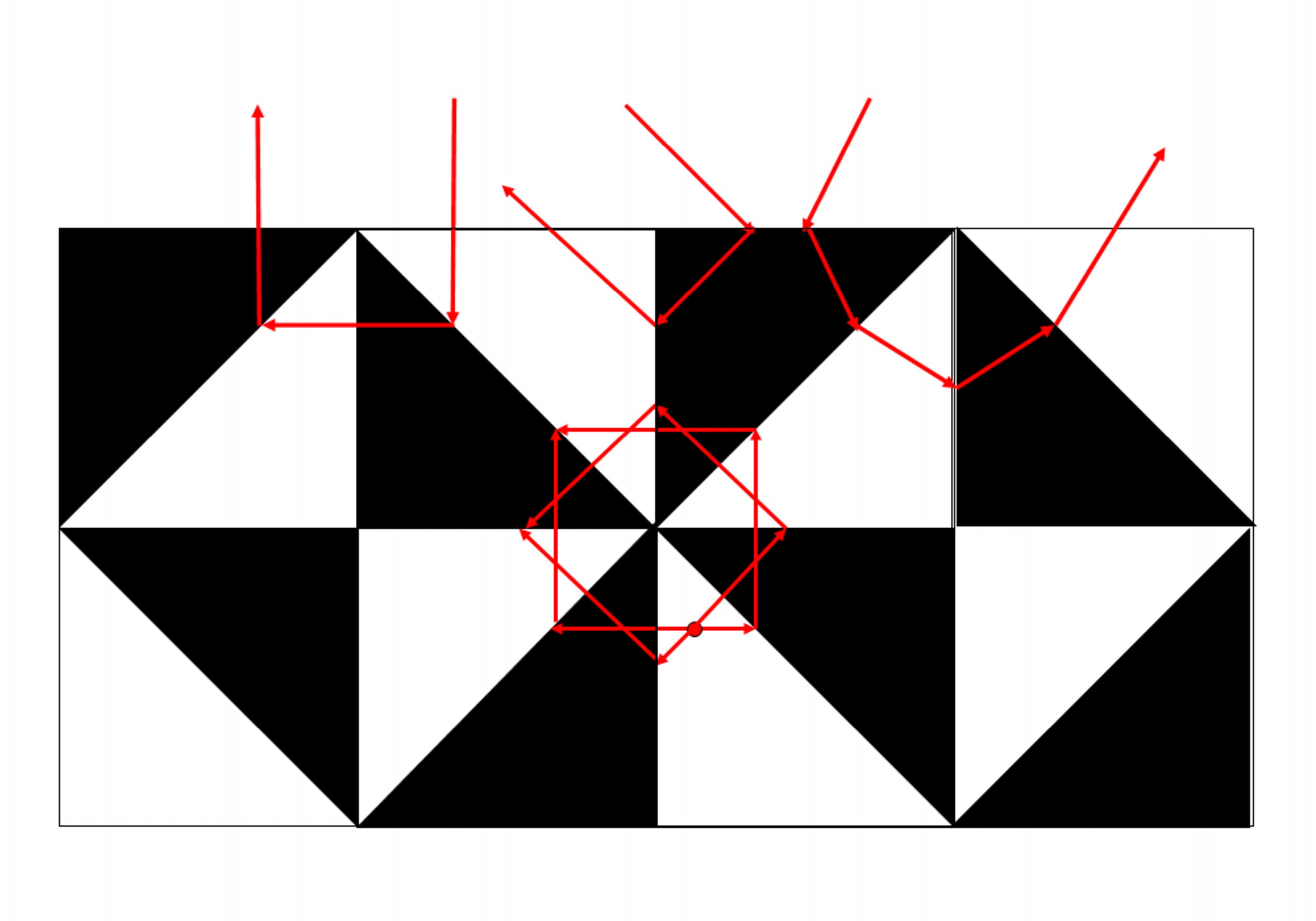}}
\caption{Periodic cell of a square checkerboard
lens with right-angled triangular inclusions sharing a corner
(clockwise). Black regions have a negative refractive index, and
white regions a positive refractive index. Any incoming ray (from
left) is reflected. Most rays emanating from a point source inside
the unit cell describe closed trajectories (giving rise to three
perfect images and four ghost images). However, the generalized lens
theorem ensures us that there is full transmission.
} \label{chess4}
\end{center}
\end{figure}

As an example of how checkerboard structures defy conventional logic,
consider a checkerboard slab lens of NRIM ($n=-1$) with square cells.
It is clear from the ray picture in Fig.~\ref{chess5} that the
rays incident from the left onto the slab will either emerge on
the other side, or will get retro-reflected depending on the initial position
and angle of the ray. This suggests that there should be partial transmittance
through this lens. The Generalized lens theorem, which is a full wave solution,
predicts that every plane wave will transmit through the system without change
in amplitude or phase. Full wave numerical solutions also show full
transmittance and this actually contradicts the ray picture, as
first reported in Ref.~\cite{sangeeta}. A more singular situation arises
in the slab lens of Fig.~\ref{chess4} with triangular checkerboards.
The ray picture predicts that every incoming ray should be reflected.
However, such a lens also displays
full transmission and zero reflectivity, which
is thus a form of extraordinary transmission (see also~\cite{pendry_contemp04} for a similar paradigm).
In both cases, the
common feature of the slab lens is the equal amount of black and
white regions, which is a prerequisite for optical space
cancellation~\cite{pendry_jpc03}.
In fact, these are examples of extraordinary
transmission mediated by excitation and scattering of surface
plasmon waves via the corners. Nevertheless, the mechanism of plasmonic guidance
involved here via the interfaces between positive and negative index
media differs substantially from the extraordinary transmission
through subwavelength holes in thick metallic films experimentally
demonstrated in~\cite{ebbesen_pendry}.

\begin{figure}[h]
\begin{center}
\hspace{2cm}\mbox{}
\scalebox{1.0}{\includegraphics[width=10cm,angle=0]{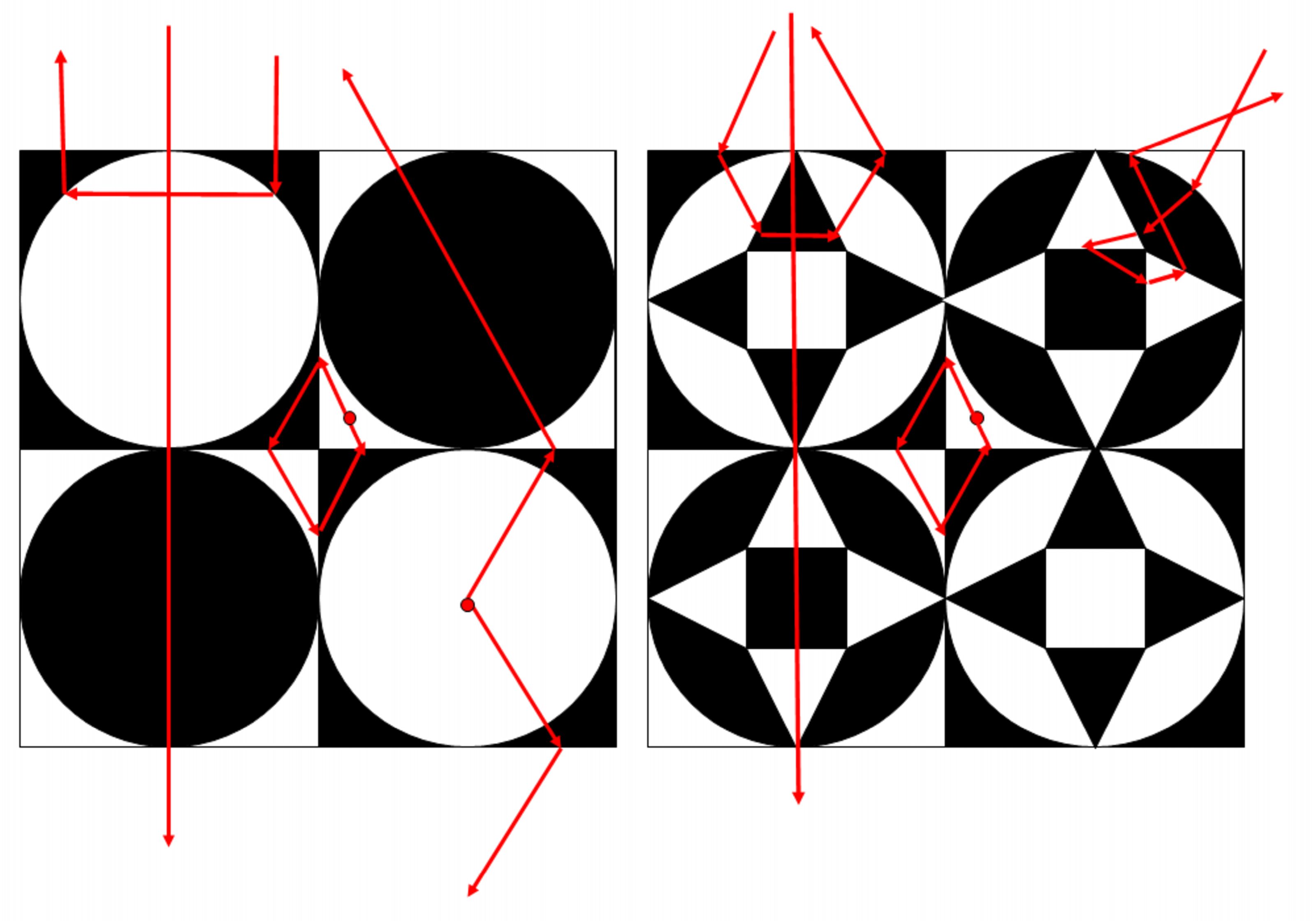}}
\caption{Left: Periodic cell of a square checkerboard
lens with circular inclusions. Black regions have a negative refractive
index, and white regions a positive refractive index. Any incoming
ray which is not orthogonal to the lens interface is reflected. Rays emanating from a point source
inside the unit cell describe closed trajectories except if it is
located at the center of circular inclusions. Right: When we further structure the lens with triangles
and squares, ray trajectories are complexified.}
\label{chess6}
\end{center}
\end{figure}

\begin{figure}[t]
\begin{center}
\hspace{0cm}\mbox{}
\scalebox{1.0}{\includegraphics[width=5cm,angle=-90]{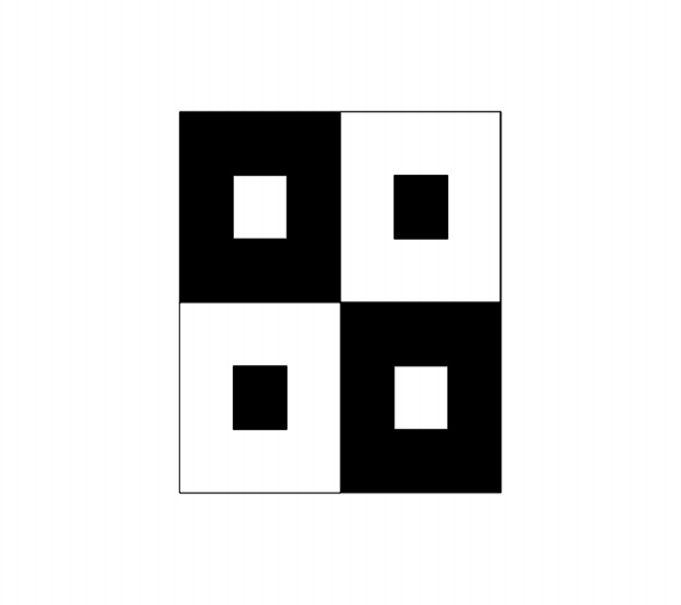}}
\scalebox{1.0}{\includegraphics[width=5cm,angle=-90]{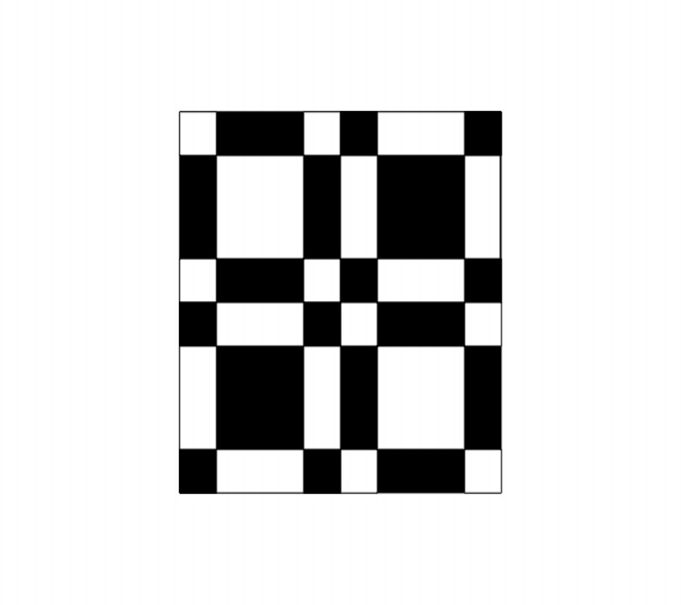}}
\scalebox{1.0}{\includegraphics[width=5cm,angle=-90]{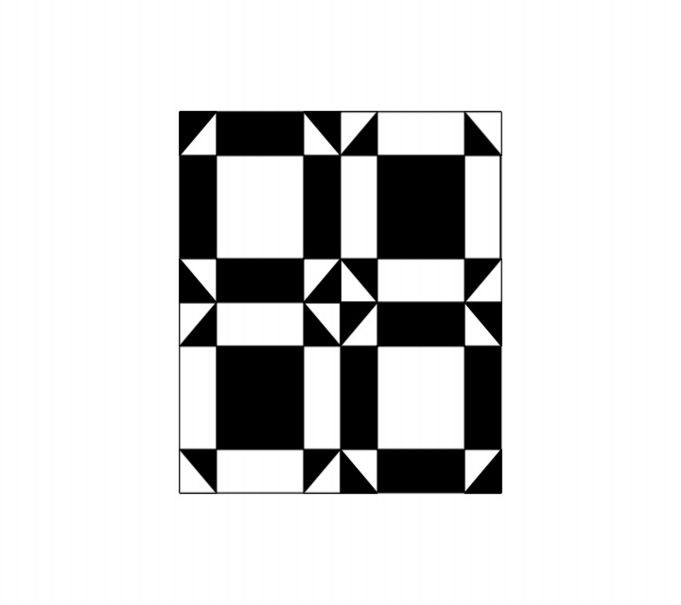}}
\caption{Left: Periodic cell of a square checkerboard
with square inclusions. Black regions are considered to have a negative
refractive index, while white regions have a positive refractive index. Square inclusions can be
small (dilute limit) or very large (thin-bridge limit).
Middle: Periodic cell of a square checkerboard
combining square and rectangular inclusions.
Right: Periodic cell of a square checkerboard
combining square, rectangular and right-angled triangular inclusions.}
\label{chess1}
\end{center}
\end{figure}

For a finite checkerboard of triangular cells with alternating
triangles having refractive index $n=\pm 1$, the ray picture
predicts that the rays  from a source placed in one of the interior
cells cannot escape from the structure if the intersection of the
wedges is completely surrounded by other points of intersection.
This suggests that such a system will very strongly confine light.
Hence the ray analyses suggest very interesting properties for
checkerboard systems. However,
as illustrated by the paradoxes of the intuitive, but approximate,
ray picture showing no transmission and the complementary theorem
showing perfect transmission, it becomes imperative to investigate
numerically the full wave solutions of finite checkerboard
structures of NRIM.

\subsection{Origami Lenses}

One might look for other ways of tiling the
plane. Nevertheless, the crystallographic restriction theorem states
that rotational symmetries in planar crystals are limited to
two-fold, three-fold, four-fold, and six-fold. Further, keeping a
balance between overall positive refractive index material and NRIM
implies that we are only left with checkerboards of either rectangular,
square or (equilateral) triangular cells. However, the unit cell can have
further structure that respects the mirror-antisymmetry conditions
as well as the crystallographic conditions. We show examples of
structured unit cells in Fig.~\ref{chess1} that have
an overall balance of positive and negative complementary media so that
the unit cell overall has zero optical pathlength. Such designs are
reminescent of Victor Vasarely's art where checkerboards have been
used to great effect~\cite{vasarely}. We term such lenses as {\it origami lenses}
 and study them principally due to curiosity. It turns out that such lenses
have very interesting electromagnetic properties as we present in the sequel.

\begin{figure}[tb]
\begin{center}
\scalebox{1.0}{\includegraphics[width=10cm,angle=-0]{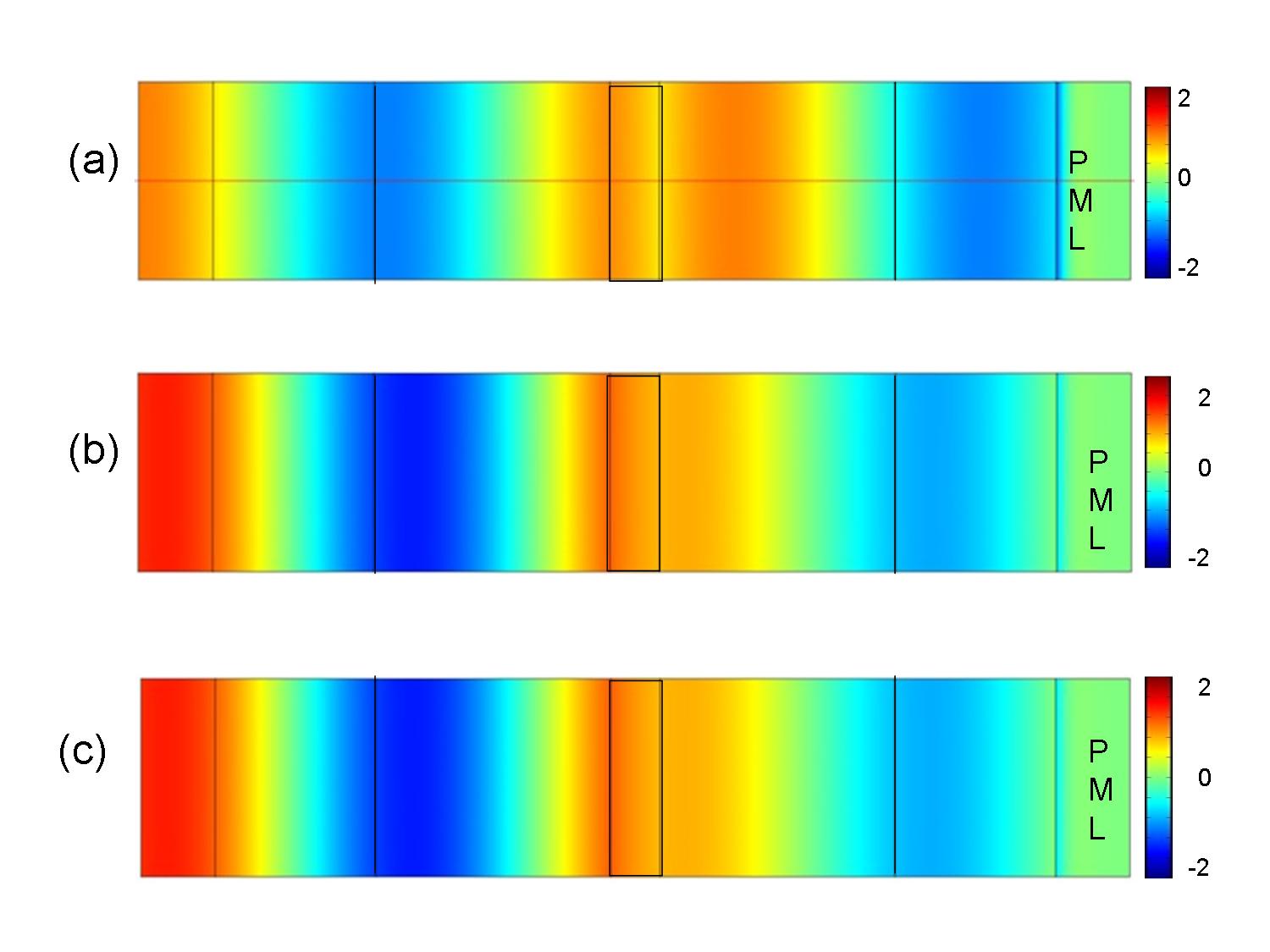}}
\caption{A rectangular slab lens with; (a)
$\varepsilon=\mu=-1$; (b) $\varepsilon=-1+\i 0.01$; (c)
$\varepsilon=-1+\i 0.1$, illustrating the transmission of a plane wave incident from the left. The results shown
are for p-polarized light and $d \simeq \lambda/10$.} \label{s5}
\end{center}
\end{figure}

\begin{figure}[tb]
\begin{center}
\scalebox{1.0}{\includegraphics[width=10cm,angle=-0]{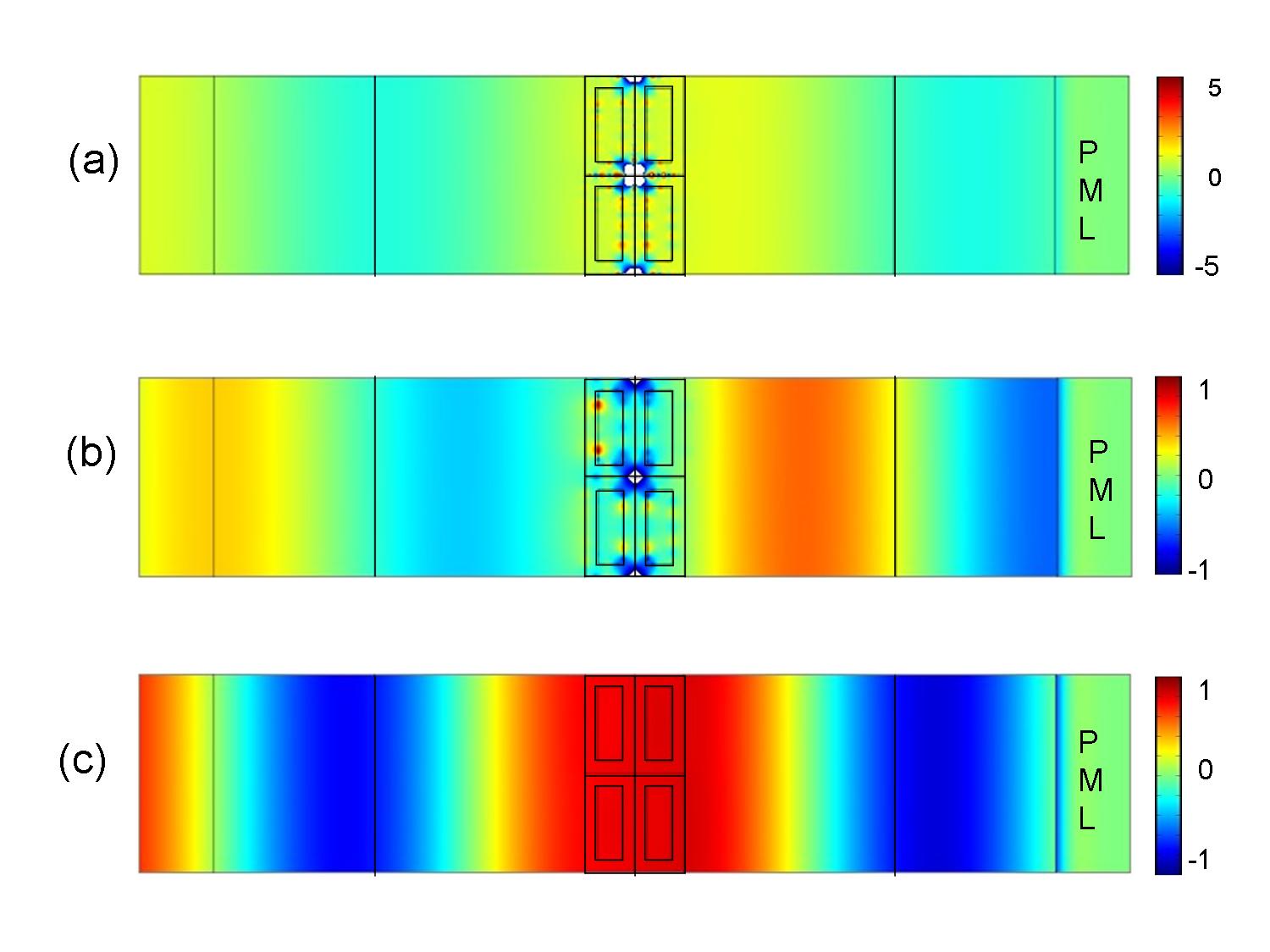}}
\caption{Origami lens with embedded rectangular cells; (a)
$\varepsilon=\mu=-1$; (b) $\varepsilon=-1+\i 0.01$; (c)
$\varepsilon=-1+\i 0.1$. A plane wave is incident from the left and transmitted across the slab. The results are for
p-polarized light and the thickness of the checkerboard $d \simeq \lambda/5$}. \label{s1}
\end{center}
\end{figure}

\begin{figure}[tb]
\begin{center}
\scalebox{1.0}{\includegraphics[width=10cm,angle=-0]{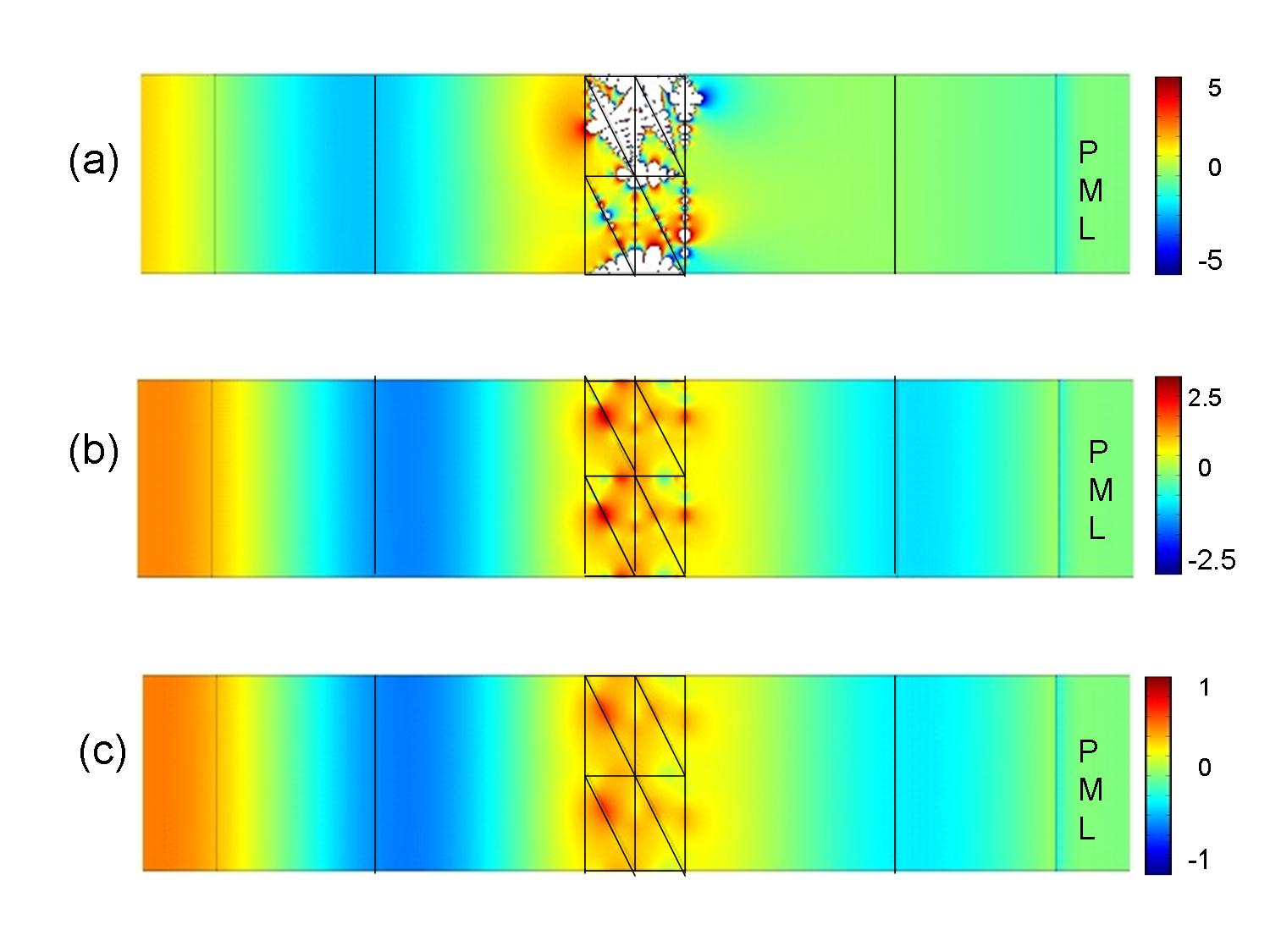}}
\caption{Origami lens with right-angled triangular cells; (a)
$\varepsilon=\mu=-1$; (b) $\varepsilon=-1+\i 0.01$; (c)
$\varepsilon=-1+\i 0.1$. As before, each of these figures illustrates the transmission of a
a plane wave incident from the left hand side, for p-polarized light across a slab whose thickness $d \simeq \lambda/5$.} \label{s2}
\end{center}
\end{figure}

\begin{figure}[tb]
\begin{center}
\scalebox{1.0}{\includegraphics[width=10cm,angle=-0]{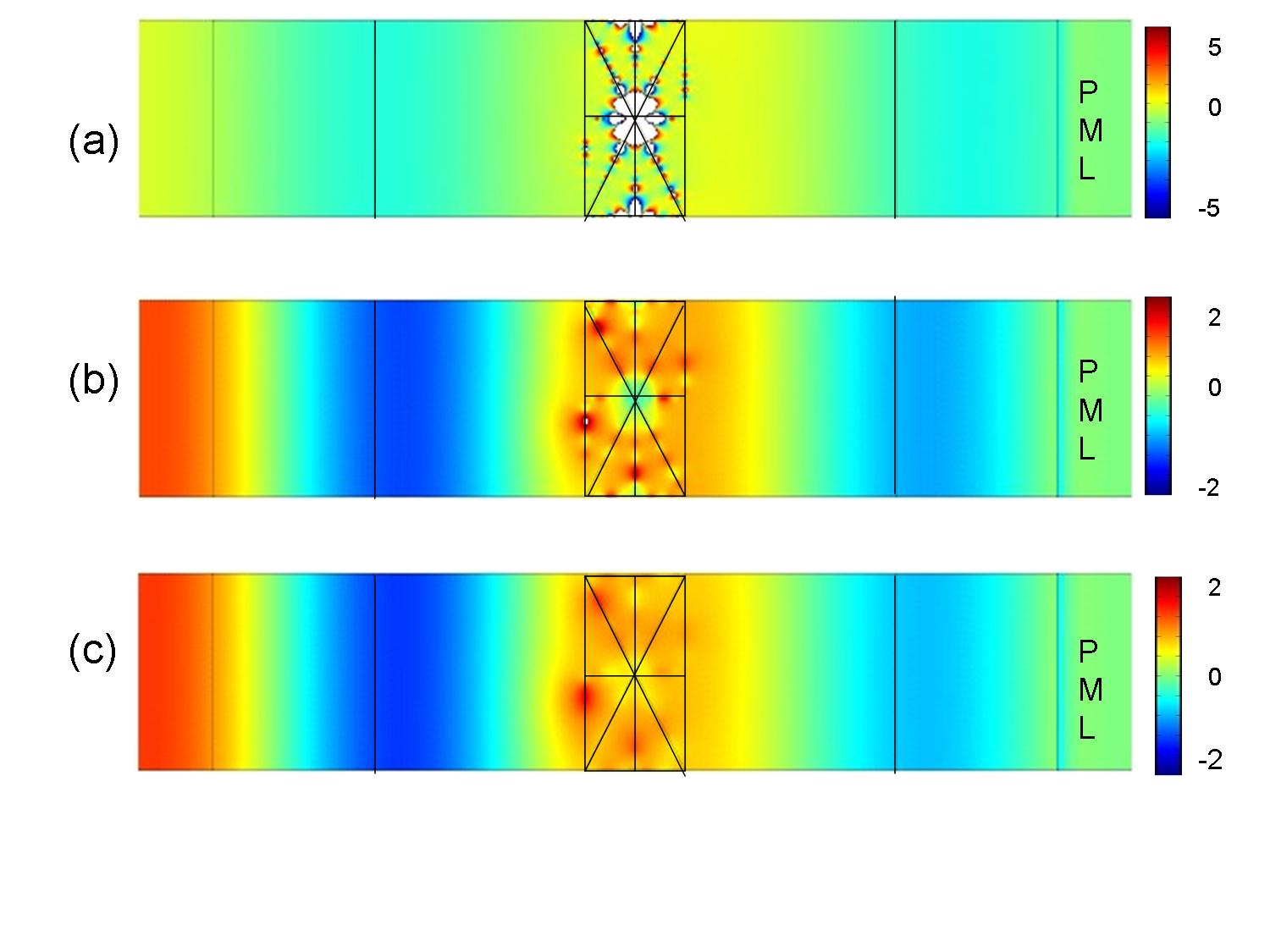}}
\caption{Origami lens with triangular cells arranged in a different pattern for (a)
$\varepsilon=\mu=-1$; (b) $\varepsilon=-1+\i 0.01$; (c)
$\varepsilon=-1+\i 0.1$. Evidently, the field distributions and the transmission, obtained for a p-polarized plane wave incident from the left, are
quite different from the ones shown in Fig.~\ref{s2}. As before, the thickness of the slab, $d \simeq \lambda/5$.} \label{s3}
\end{center}
\end{figure}

\begin{figure}[tb]
\begin{center}
\scalebox{1.0}{\includegraphics[width=10cm,angle=-0]{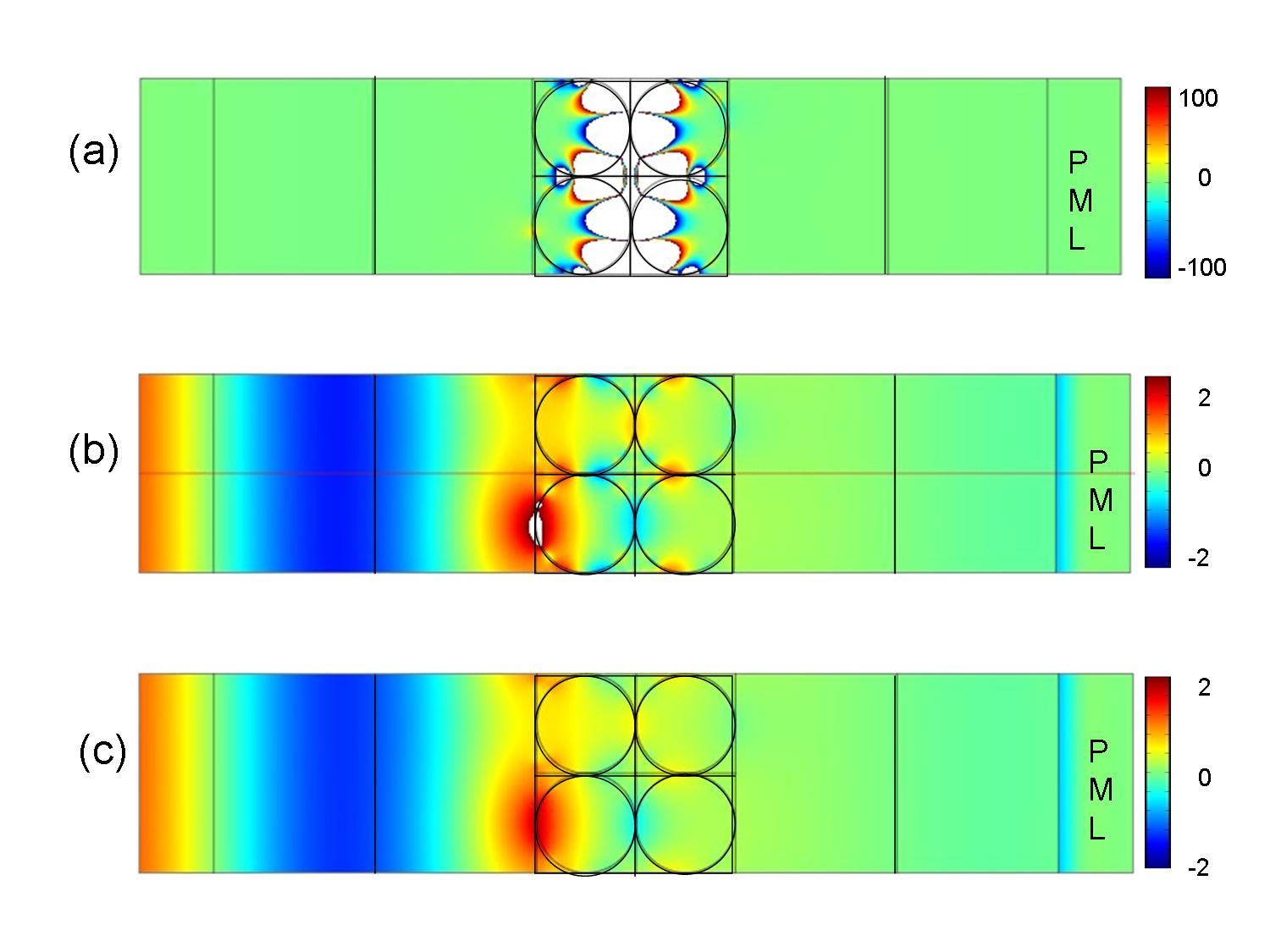}}
\caption{Origami lens with embedded circular cells for (a)
$\varepsilon=\mu=-1$; (b) $\varepsilon=-1+\i 0.01$; (c)
$\varepsilon=-1+\i 0.1$. The transmission and the field distributions for a p-polarized plane wave incident from the left on a checkerboard slab $d \simeq \lambda/5 $(as in the preceding figures
) have been shown.} \label{s4}
\end{center}
\end{figure}

We report in Fig.~\ref{s5} the finite element computations for a
planewave incident on a slab lens of infinite extent along the
transverse direction. P-polarized radiation is incident from the left and $d$, the width of the slab,
is $\lambda/10$. In Fig.~\ref{s5}, the transmission properties of three types of slab lens has been illustrated: (a) the perfect
slab lens and dissipative slab lenses with
(b) $\epsilon = -1 + i 0.01, \mu = 1$ and (c) $\epsilon = -1 + i 0.1, \mu = 1$.
Perfect transmission is obtained in Fig.~\ref{s5}(a) and this is seen to decrease in Fig.~\ref{s5}(b) and (c).
These results have been summarized in Table~\ref{tab1}.
We then add some complementary media within
the slab lens, see Fig.~\ref{s1}-\ref{s4}. We note that while the
electromagnetic is clearly enhanced within the origami lenses, the
planewave still goes unperturbed as the overall amount of positive
and negative index media is well balanced.

\subsection{Infinite Checkerboards and transformational optics}

We have seen that the electromagnetic paradigm of negative refraction is the perfect lens, whereby a homogeneous
slab of negative refractive index material maps the image plane onto the source plane, which
has been recently revisited through an optical space folding approach~\cite{crp2009}. In fact,
any heterogeneous anisotropic medium satisfying certain anti-symmetry properties fulfils this
optical space folding (leading to a cancelation of the optical path)~\cite{pendry_jpc03}. It is
thus possible to design very complex metamaterials using a very simple group theoretical approach~\cite{guenneau_njp05},
and this already led to the discovery of infinite checkerboards alternating
cells of complementary media. However, a systematic study of such checkerboards is far from being
easy, as standard numerical packages such as finite elements might fail to converge in the analysis of such
strongly resonant metamaterials, as shown in~\cite{sangeeta}. Some rigorous mathematical study of
such sign-shifting media has been performed by the team of Anne-Sophie Bonnet-Bendhia~\cite{sophie}.
It is shown there that the Lax-Milgram lemma does not apply anymore (lack of ellipticity), but it is
still possible to invoke the Fredholm alternative in order to be assured that any numerical solution
found by a finite element algorithm satisfying the prerequisite transmission conditions on the interfaces
between complementary media (tangential `anti-continuity' of the field) and adhoc outgoing wave conditions
at infinity (such as Sommerfeld ones) exists in its own right (is not spurious). It was also observed in~\cite{sangeeta}
that the PHOTON code is superior to the standard commercial package COMSOL\texttrademark in handling transmission type problems
through rectangular checkerboard lenses with or without dissipation. In the next section, we explore more intricate
checkerboard lenses displaying for instance some thin bridges between complementary media within which
the electromagnetic field is further enhanced.
Some paradoxes occur when the checkerboard lens becomes infinite in all-space dimensions:
in this case, one has to think of the periodic structure as being born of a torus and any source located within a unit
cell will show an infinite number of images in any other cell of the checkerboard (which actually reduces to one cell).

Transformational optics is a very powerful tool enhancing creative
thinking in the context of metamaterials: The first step is to
define a map from the curvilinear metric we wish to create (keeping
in mind that light will follow the geodesics) onto our usual
Euclidean metric (within which geodesics happen to be straight
lines).

\noindent In the context of generalized lenses, we want to fold the
optical space back onto itself, and this leads to negative
coefficients within the permittivity and permeability matrices. The
coordinate transformation is given by
\begin{equation}
\hspace{-2cm}\mbox{}
\left\lbrace
\begin{array}{ll}
x'_1 & = x_1 \; , \nonumber \\
x'_2 & = x_2 \; , \nonumber \\
x'_3 & = x_3 -d \; , \hbox{ if } x'_3 < d/2 \hbox{, or } -x_3 \hbox{
if } -d/2 < x'_3 < d/2 \hbox{, or } x_3 +d \hbox{ if } d/2 < x'_3
\end{array}
\right. \label{transfolens1d}
\end{equation}
where d is the thickness of the lens.

This change of co-ordinates is characterized by the Jacobian of the
transformation:

\begin{equation}
\hspace{-2cm}\mbox{}
\begin{array}{ll}
\mathbf{J}_{{\bf xu}} = \displaystyle{ \large
\frac{\partial(x_1,x_2,x_3)}{\partial(u_1,u_2,u_3)} } =

\left(
\begin{array}{ccc}
\frac{\partial x_1}{\partial u_1}  &\frac{\partial x_1}{\partial u_2}   &\frac{\partial x_1}{\partial u_3}\\
\frac{\partial x_2}{\partial u_1}  &\frac{\partial x_2}{\partial u_2}   &\frac{\partial x_2}{\partial u_3}\\
\frac{\partial x_3}{\partial u_1}  &\frac{\partial x_3}{\partial u_2}   &\frac{\partial x_3}{\partial u_3}\\
\end{array}
\right) \; ,
\end{array}
\hbox{with}
\left(%
\begin{array}{c}
  dx_1 \\
  dx_2 \\
  dx_3 \\
\end{array}%
\right) = \mathbf{J}_{\bf xu}
\left(%
\begin{array}{c}
  du_1 \\
  du_2 \\
  du_3 \\
\end{array}%
\right) \; ,
\end{equation}

The second step is to link the Jacobian of these two metrics to
constitutive parameters within the Maxwell system: the curvilinear
metric is described by anisotropic heterogeneous permittivity and
permeability matrices, while the Euclidean one is associated with
identity matrices. Indeed, the inhomogeneous anisotropic
metamaterial designed as a result of the co-ordinate transform is
described by a transformation matrix ${\bf T}$ (metric tensor) via:
\begin{equation}
\underline{\underline{\varepsilon'}} =\varepsilon \mathbf{T}^{-1} \;
, \underline{\underline{\mu'}}=\mu \mathbf{T}^{-1} \; , \mathbf{T} =
\frac{\mathbf{J}^T \mathbf{J}}{det(\mathbf{J})} \; .
\end{equation}

The coordinate transform (\ref{transfolens1d})leads to the identity
for the transformation matrix ${\bf T}$ outside the lens, whereas
inside the lens i.e. for $-d/2 < x'_3 < d/2$, $\partial x_3/\partial
x'_3=-1$ which flips the sign of ${T}_{33}$, so that the material
properties differ from free space only in the $x_3=x'_3$ direction,
whereby the scalar permittivity $\varepsilon=\varepsilon_1$ and the
transformed (scalar) permittivity $\varepsilon'=\varepsilon_2$, and
similarly for the permeability.

However, in the case of anisotropic
$\underline{\underline{\varepsilon}}$ and
$\underline{\underline{\mu}}$, the transformed medium is now
characterized by
\begin{equation}
\hspace{-2cm}\mbox{}
\begin{array}{lll}
&\underline{\underline{\varepsilon'}} =\rm{det}(\mathbf{J})
\left(\mathbf{J}^{-1}\underline{\underline{\varepsilon}}\mathbf{J}^{-T}\right)
= (-1)\left(
\begin{array}{rrr}
        1 & 0 & 0 \\
        0 & 1 & 0 \\
        0 & 0 &-1
         \end{array} \right)
\left( \begin{array}{lll}
        \varepsilon_{11} & \varepsilon_{12} & \varepsilon_{13} \\
        \varepsilon_{21} & \varepsilon_{22} & \varepsilon_{23} \\
        \varepsilon_{31} & \varepsilon_{32} & \varepsilon_{33}
         \end{array} \right)
\left(
\begin{array}{rrr}
        1 & 0 & 0 \\
        0 & 1 & 0 \\
        0 & 0 & -1
         \end{array} \right)
\\
&= \left( \begin{array}{lll}
        -\varepsilon_{11} & -\varepsilon_{12} & +\varepsilon_{13} \\
        -\varepsilon_{21} & -\varepsilon_{22} & +\varepsilon_{23} \\
        +\varepsilon_{31} & +\varepsilon_{32} & -\varepsilon_{33}
         \end{array} \right),
         \\
&\underline{\underline{\mu'}} =\rm{det}(\mathbf{J})
\left(\mathbf{J}^{-1}\underline{\underline{\mu}}\mathbf{J}^{-T}\right)
= (-1)\left(
\begin{array}{rrr}
        1 & 0 & 0 \\
        0 & 1 & 0 \\
        0 & 0 &-1
         \end{array} \right)
\left( \begin{array}{lll}
        \mu_{11} & \mu_{12} & \mu_{13} \\
        \mu_{21} & \mu_{22} & \mu_{23} \\
        \mu_{31} & \mu_{32} & \mu_{33}
         \end{array} \right)
\left(
\begin{array}{rrr}
        1 & 0 & 0 \\
        0 & 1 & 0 \\
        0 & 0 & -1
         \end{array} \right)
\\
&=
\left( \begin{array}{lll}
        -\mu_{11} & -\mu_{12} & +\mu_{13} \\
        -\mu_{21} & -\mu_{22} & +\mu_{23} \\
        +\mu_{31} & +\mu_{32} & -\mu_{33}
         \end{array} \right),
\end{array}
\end{equation}
which is in accordance with (\ref{sarlens2}).

We note that there is no change in the impedance of the media, since
the permittivity and permeability undergo the same geometric
transformation: the perfect lens is impedance-matched with its
surrounding medium (air, say) so that no reflection will occur at
its interfaces.

Altogether, if we consider a complex medium described by general
dielectric permittivity and magnetic permeability tensors given by
\begin{equation}
\underline{\underline{\varepsilon_{1}}} = \left( \begin{array}{lll}
        \varepsilon_{11} & \varepsilon_{12} & \varepsilon_{13} \\
        \varepsilon_{21} & \varepsilon_{22} & \varepsilon_{23} \\
        \varepsilon_{31} & \varepsilon_{32} & \varepsilon_{33}
         \end{array} \right), ~~~~~
\underline{\underline{\mu_{1}}} = \left( \begin{array}{lll}
        \mu_{11} & \mu_{12} & \mu_{13} \\
        \mu_{21} & \mu_{22} & \mu_{23} \\
        \mu_{31} & \mu_{32} & \mu_{33}
         \end{array}  \right) \; , \; -d < x_3 <
        0 \; ,
\label{sarlens1}
\end{equation}
then the resulting complementary medium is given by
\begin{equation}
\hspace{-2cm}\mbox{}
\underline{\underline{\varepsilon_{2}}} = \left( \begin{array}{lll}
        -\varepsilon_{11} & -\varepsilon_{12} & +\varepsilon_{13} \\
        -\varepsilon_{21} & -\varepsilon_{22} & +\varepsilon_{23} \\
        +\varepsilon_{31} & +\varepsilon_{32} & -\varepsilon_{33}
         \end{array} \right), ~~~
\underline{\underline{\mu_{2}}} = \left( \begin{array}{lll}
        -\mu_{11} & -\mu_{12} & +\mu_{13} \\
        -\mu_{21} & -\mu_{22} & +\mu_{23} \\
        +\mu_{31} & +\mu_{32} & -\mu_{33}
         \end{array} \right) \; , \; 0 < x_3 < d \; ,
\label{sarlens2}
\end{equation}
which is the result first derived in ~\cite{pendry_jpc03} and
retrieved using group theory (symmetries of Maxwell's equations) in
~\cite{guenneau_njp05}. The entries in
$\underline{\underline{\varepsilon}}$ and
$\underline{\underline{\mu}}$ can also be spatially varying along
$x$ and $y$. This covers the case of perfect corner reflectors of
$2n$-fold skew-symmetry and we are therefore ensured of the
cancelation of the optical path. It is worth noting that the
generalized lens theorem was also applied to infinite checkerboards
of skew-symmetry in ~\cite{guenneau_njp05}.

\subsection{The anisotropic lens}

\begin{figure}[tbp]
\begin{center}
\scalebox{1.0}{\includegraphics[width=10cm,angle=-0]{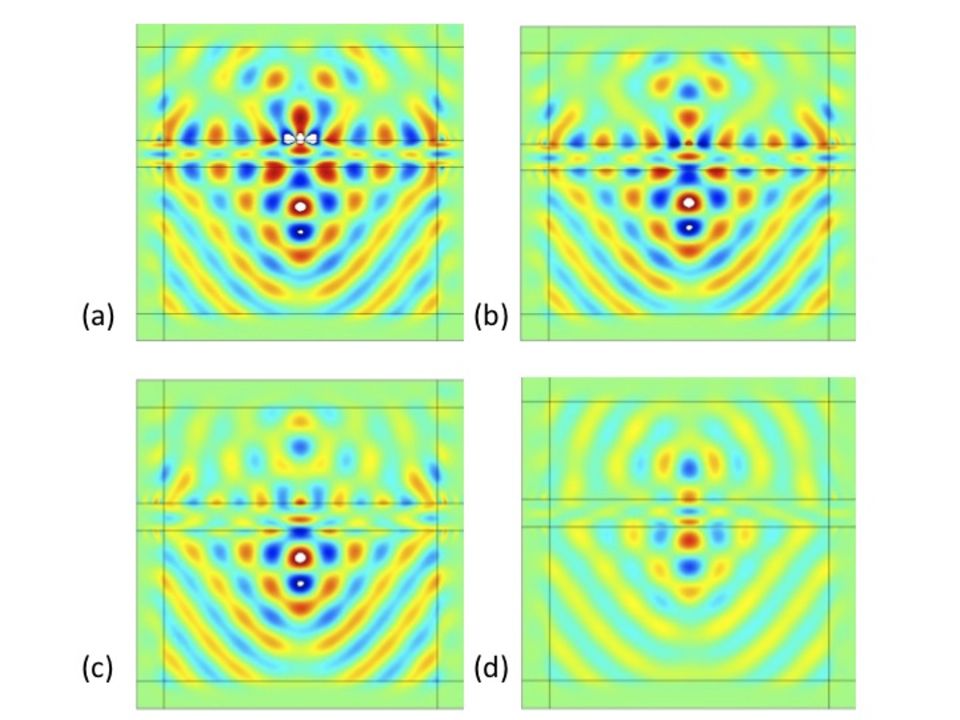}}
\caption{Anisotropic slab lens functioning as a generalized version of the lens described in Ref.~\cite{kafesaki_prb09}.
The imaging properties of an anisotropic slab lens of thickness $d$ whose permittivity is given by $\epsilon_{2x} = -2-\cos y + 0.1*i $, $\epsilon_{2x} = 1/(-2-\cos y) + 0.1*i$ and $ \mu_{2y} = -2-\cos y + 0.1*i $, for various positions of the source. (a) The source is at the slab boundary, (b) The source is at a distance $d/4$ from the slab, (c) The source is at $d/2$, (d) The source is at a distance $d$ from the slab.
} \label{anisolens}
\end{center}
\end{figure}

In this section, we present the focussing properties of a single slab lens of anisotropic
material. This is a generalization of the anisotropic `far-field superlens' proposed by Shen
{\it et. al.} in Ref.~\cite{kafesaki_prb09}. The slab lens is impedance-matched to the surrounding
medium (vacuum), thus eliminating the possibility of reflection at the interfaces and satisfies
the following conditions specified in ~\cite{kafesaki_prb09}:
\begin{equation}
\frac{\epsilon_{2x}}{\mu_{2y}} = \frac{\epsilon_1}{\mu_1}; ~~\epsilon_{2x}\epsilon_{2z} = \epsilon_1^2.
\end{equation}
Here, $\epsilon_{2x}$, $\epsilon_{2z}$ and $\mu_{2y}$ are the diagonal components of the permittivity and the permeability tensors (defined
to be diagonal matrices.) The subscripts $1,2$ refer to the surrounding medium (vacuum) and the anisotropic slab, respectively. We choose the following
general form for a spatially varying permittivity : $\epsilon_{2x} = -2-\cos y + 0.01*i $, $\epsilon_{2x} = 1/(-2-\cos y) + 0.01*i$ and
$ \mu_{2y} = -2-\cos y + 0.01*i $. Fig.~\ref{anisolens} illustrates the focussing action of this slab lens. We find that such a slab lens with generalized spatially varying parameters exhibits focussing properties and is thus, a more generalized version of the anisotropic slab lens discussed by
Kafesaki {\it et. al}.

\begin{figure}[tbp]
\begin{center}
\scalebox{1.0}{\includegraphics[width=14cm,angle=-0]{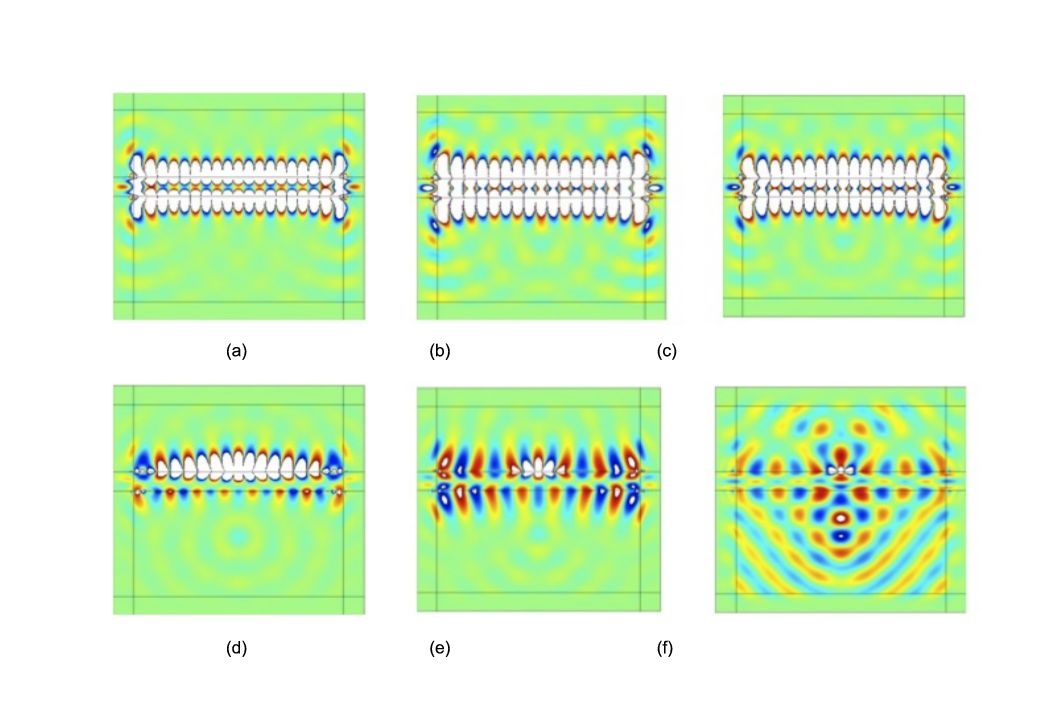}}
\caption{Variation of the number of spatial oscillations of surface plasmons at the anisotropic slab lens interface: The absorption in the slab is increased from $10^{-6}$ to $0.1$, from (a) to (f). It is noted that the variation is similar to Fig. \ref{logplots}.} \label{anisolensa}
\end{center}
\end{figure}

We have also investigated the dependence of the properties of this generalized slab lens on the dissipation in the slab. The dissipation affects the spatial oscillation of the surface plasmons at the interfaces, in the same manner as described in Section~\ref{genl}. This has been illustrated in Fig.~\ref{anisolensa}, where the point source is located on the slab boundary.

\subsection{Three dimensional checkerboards of complementary media versus Four Color Theorem}
\begin{figure}[t]
\begin{center}
\scalebox{2.0}{\includegraphics[width=5cm,angle=0]{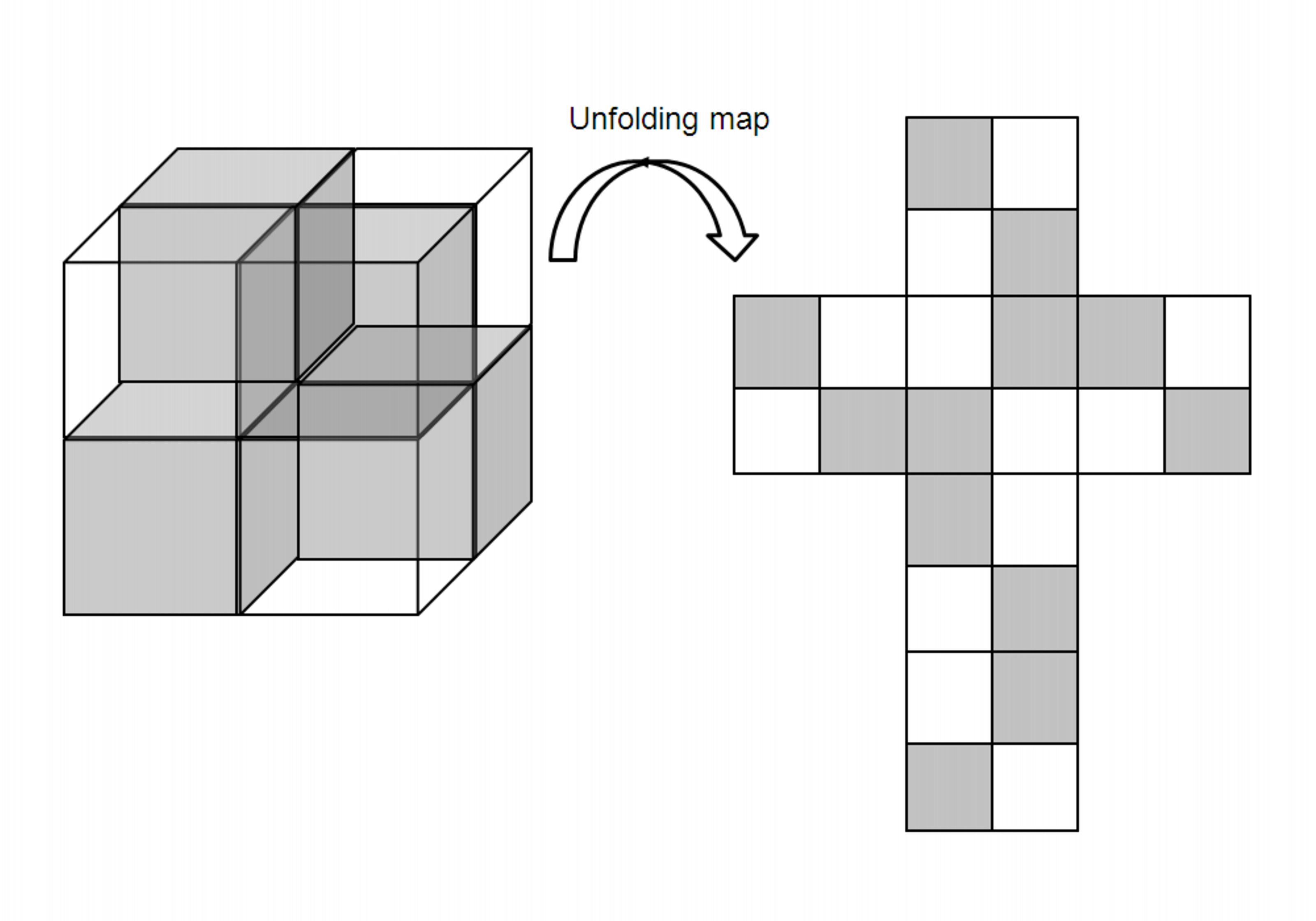}}
\vspace{0cm}\mbox{}
\caption{Left: Unit cell for a three-dimensional two-phase periodic checkerboard consisting of
cubes filled with positive (white) and negative (grey) refractive index media.
Right: unfolded checkerboard exemplifying the role of symmetries and rotations
in the design.
Interfaces between
cubes (i.e. faces and edges) support a host of surface and edge plasmons.}
\label{origami1}
\end{center}
\end{figure}

\begin{figure}[t]
\begin{center}
\scalebox{2.0}{\includegraphics[width=5cm,angle=0]{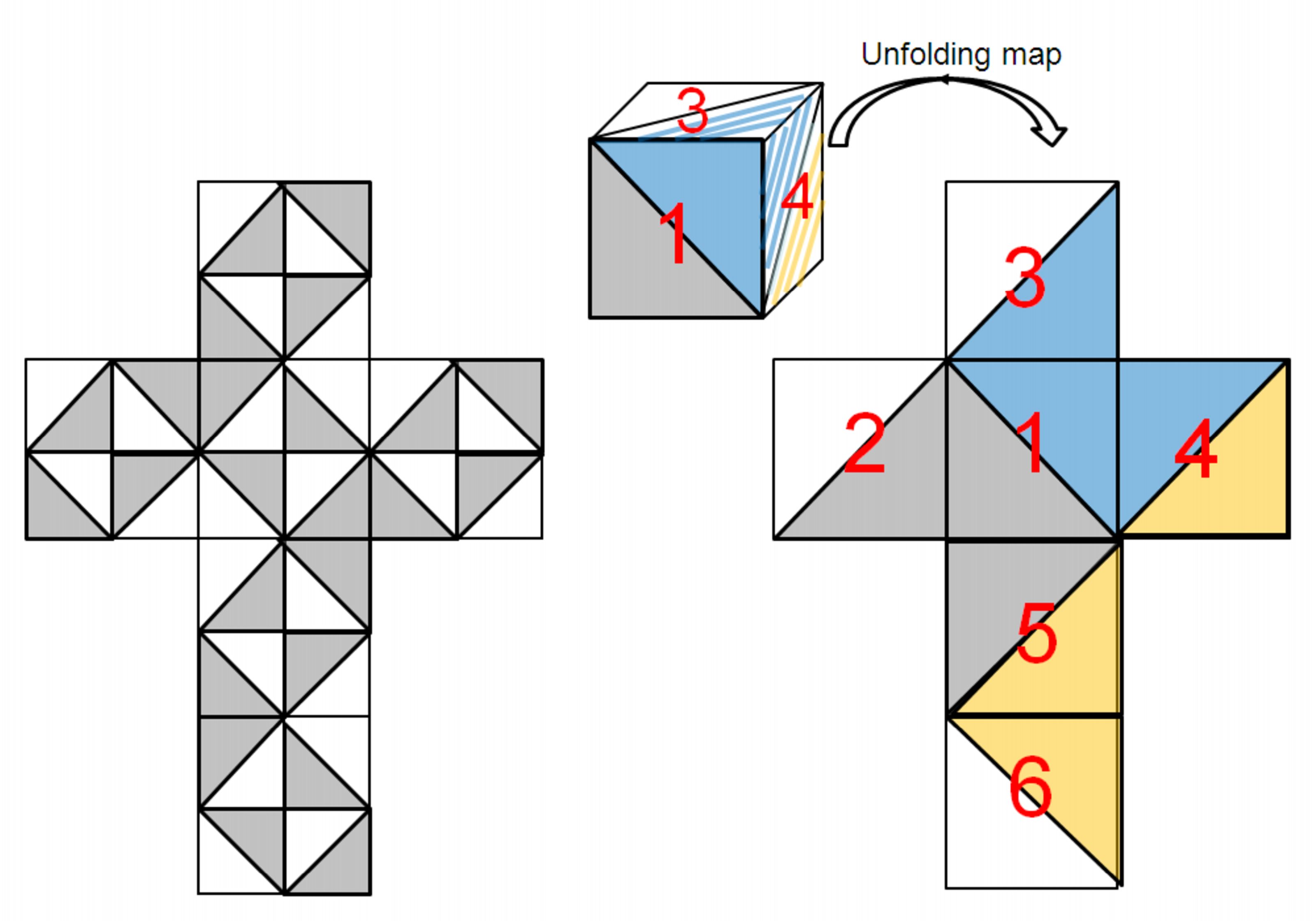}}
\vspace{0cm}\mbox{}
\caption{Left: Unfolded checkerboard consisting of right-angled tetrahedra filled
with positive (white) and negative (grey) refractive index media. This leads to an
impossible three-dimensional checkerboard.
Right: Elementary brick of a four phase three-dimensional checkerboard (see Fig.~\ref{origami3}) with
its unfolded counterpart consisting of four complementary phases.
These negative and positive results are consistent with the four color theorem.}
\label{origami2}
\end{center}
\end{figure}

\begin{figure}[t]
\begin{center}
\scalebox{2.0}{\includegraphics[width=5cm,angle=0]{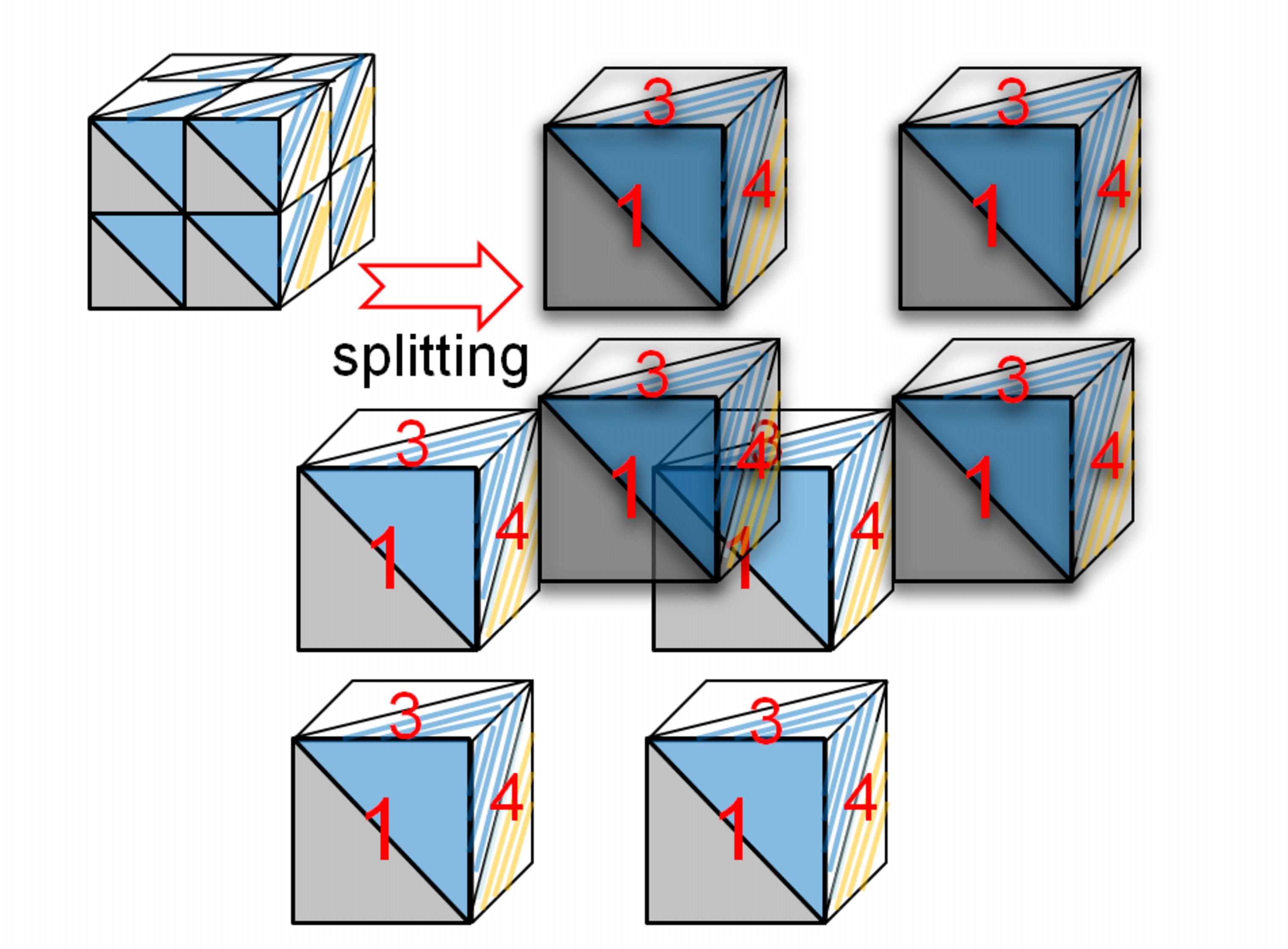}}
\vspace{0cm}\mbox{}
\caption{Four phase three-dimensional checkerboard consisting of right-angled tetrahedra filled
with positive (white, $n=n_1$, blue $n=n_2$) and negative (grey, $n=-n_1$, yellow, $n=-n_2$)
refractive index media.}
\label{origami3}
\end{center}
\end{figure}

It was shown in Ref.~\cite{guenneau_njp05} that it is possible to fold the Euclidean
space back onto itself using an alternation of positively and negatively refracting cubic regions.
We reproduce in Fig.~\ref{origami1} the original idea of Guenneau, Vutha  and Ramakrishna.
However, these authors did not consider the case of 3D origami checkerboards, such as shown
in Figs.~ \ref{origami2} and ~\ref{origami3}: Such checkerboards are more challenging as their
design is constrained by the four color theorem.

In mathematics, the famous four color theorem, or the four color map theorem, established in 1976 by Kenneth Appel and
Wolfgang Haken~\cite{appel1,appel2} using computational techniques, states that:

\underline{\bf Theorem}
Given any separation of a plane into contiguous regions, producing a figure called a map, no more than four colors are required to color the regions of the map so that no two adjacent regions have the same color. Two regions are called adjacent only if they share a border segment, not just a point.

In the present case, this theorem warns us that a two-phase three-dimensional checkerboard with intricated pattern
might not be possible at all as its design requires to work with unfolded regions in the plane which should not share
frontiers, just like in the four color theorem. We are indeed enable to propose any two-phase checkerboard consisting
of right-angled tetrahedra, as shown in Fig.~\ref{origami2} and ~\ref{origami3}. This is to the best of our knowledge
the first example of an impossible three-dimensional checkerboard in the context of plasmonics.

\section{Numerical analysis of checkerboard structures}

We now examine the response of a NRIM checkerboard lattice lens with finite
transverse size, when its unit cells exhibit a four-fold geometry with
complex patterns. We note that the case of thin-bridges inclusions within the
checkerboard structures  represents a singular situation
which is very hard to handle with the finite element method.

\subsection{Transfer Matrix Method Calculations}

The response of some novel checkerboards was calculated using the PHOTON codes based on the Transfer Matrix Method.
A schematic representation of such checkerboards has been shown in Fig.~\ref{chess1}, extreme left. These calculations
are very sensitive to numerical errors. Accurate calculations for such highly singular structures requires a very fine numerical grid. In addition,
the regions of opposite index media must have equal thicknesses in order to be optically complementary and satisfy the Generalized
Lens theorem. The use of an unsatisfactory numerical grid results in the appearance of numerical artifacts, in the form of
resonances at wave vectors $k_x/k_0 \sim 1$, whereas no resonances are predicted to occur. We have dealt with this issue in Ref.~\cite{sangeeta} and
shown that an optimized numerical grid pushes this spurious resonance to $k_x \simeq 3 k_0$. If the thicknesses  of the adjacent
regions of the checkerboards differs even by a small amount, spurious resonances crop up again. The importance of numerical
accuracy has been discussed in Ref.~\cite{sangeeta}. Similar effects are observed if a less accurate grid (consisting of 202 points) is used or if
the adjacent cells of the checkerboard differ by a few nm.

\begin{figure}[tbp]
\hspace{1cm}\mbox{}
\includegraphics[angle = -0, width = 1.0\columnwidth]{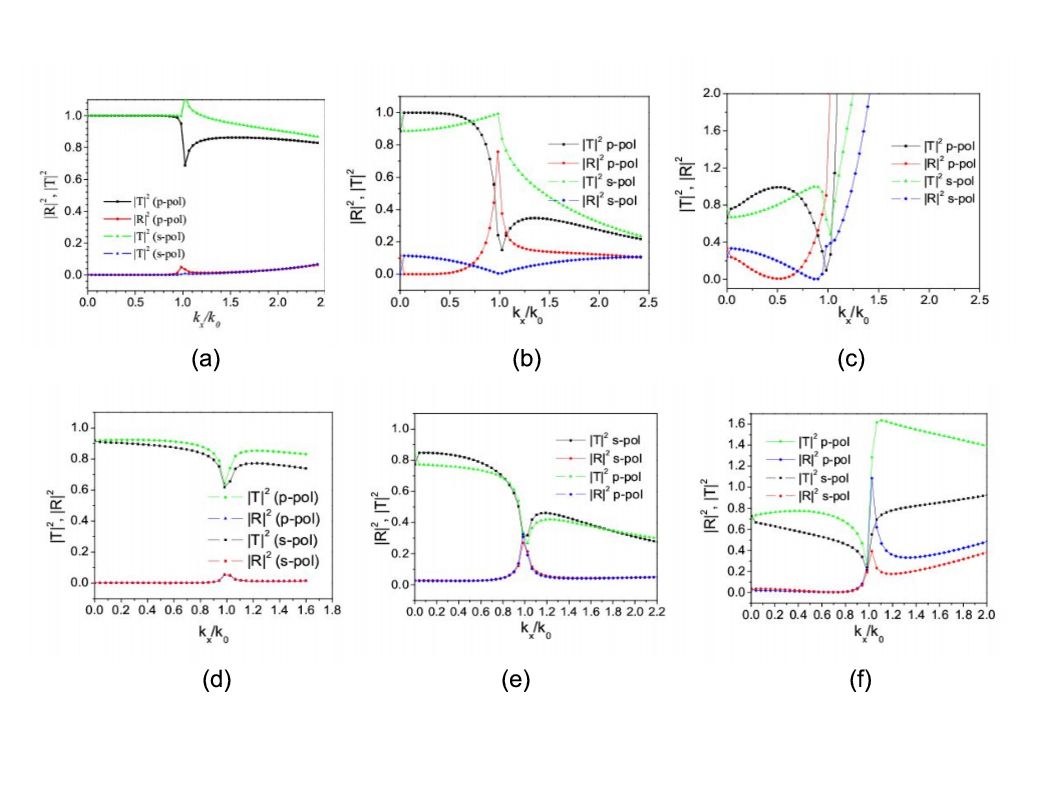}
\caption{\label{graphs} Transfer Matrix analysis of the checkerboard structures shown in Fig.~\ref{chess1} (left) illustrating their transmission properties for s- and p-polarizations. The transmission properties change as the thickness of the embedded regions changes. As the embedded regions become smaller in size, (for the graphs from left to right, top and bottom panels) the transfer matrix calculations begin to show considerable deviations. (a,d) represent the transmission properties of non-dissipative and dissipative checkerboards, respectively, whose embedded regions are fairly large in size. (b,e) illustrate the transmittive properties of non-dissipative and dissipative checkerboards, respectively, whose embedded regions are of intermediate size. (c,f) illustrate the same properties for non-dissipative and dissipative checkerboards with extremely small embedded regions.}
\end{figure}

In the upper panel of Fig.~\ref{graphs}, we present the result of our transfer matrix calculations for an optimized grid consisting of
262 points along $\hat{x}$ and 106 points along $\hat{z}$. These media exhibit nearly
unit transmittivity and zero reflectivity even for sub-wavelength wave vectors. However, when the embedded regions are very small in size, there appears to be significant deviations from the complementary lens theorem. But this is more likely to be a constraint due to the grid used, rather than implying any actual deviation of the physical behaviour of the system from the complementary behaviour. The transmittivity indicates that the system behaves as predicted by the Generalized Lens theorem. As seen in the earlier cases, the accuracy of the numerical grid is very important. Any small mismatches in the widths of the regions results in the appearance of numerical artifacts. Otherwise, unit transmittivity and zero reflectivity are obtained.

\subsection{Finite element analysis of dissipative and non-dissipative finite checkerboards}

We numerically show in Fig.~\ref{s1} to Fig.~\ref{s4} , the response of checkerboard systems
consisting of cells alternating air and NRM with increasing dissipation
for a planewave incident from the left placed on the checkerboard
for the P-polarization. In all these cases studied, the width of the checkerboard $d \simeq \lambda/10$.
We solve the Maxwell system using Finite Edge Elements (also known as Whitney forms) which naturally fulfill
transmission conditions for the tangential components of the electromagnetic field at interfaces between positive and negative
index media (hence exhibiting two anti-parallel wave-vectors at both sides of such interfaces). Also, outgoing wave conditions ensuring
well-posedness of the problem (existence and uniqueness of the solution) are enforced through implementation of Berenger Perfectly
Matched Layers within the rightmost rectangular domain ~\cite{pml}. Moreover, the expression of the incident plane wave is enforced on the leftmost
boundary, the field is set to be zero on the rightmost boundary, and periodicity conditions are in order on the top and bottom walls. Two vertical
lines located halfway from the checkerboard lens and leftmost and rightmost boundary conditions are used to compute the transmission through the line
integration of the value of the scattered field. This leads to the following numerical values (to be compared to transfer matrix results where possible):

\begin{table}
  {\begin{tabular}{|c|c|c|}
      \hline

NRM checkerboard lens&      $\varepsilon=\mu=-1$   &$\varepsilon=-1+\i*0.01$ \\
\hline

Slab lens &$100$&$99.3$  \\ \hline%

Embedded rectangles $(\square)$ (Fig.~\ref{chess1},~Left) &$100$ &$82.3$  \\ \hline%

Triangular$(\bigtriangleup)$ (Fig.~\ref{s3}) &$100$&$77.5$     \\ \hline%

Triangular$(\bigtriangledown)$ (Fig.~\ref{s2}) &$100.1$&$67.7$     \\ \hline%

Embedded circle (~~\circle{6})(Fig.~\ref{chess6}) &$100.2$ &$4.963$  \\ \hline%

\hline
\hline

NRM checkerboard lens & $\varepsilon=-1+\i*0.1$ & $\varepsilon=-1+\i*0.4$ \\
\hline

Slab lens  &$92.5$ &  $75.4$  \\ \hline%

Embedded rectangles $(\square)$ (Fig.~\ref{chess1},~Left)  & $69.9$ & $55.3$   \\ \hline%

Triangular$(\bigtriangleup)$ (Fig.~\ref{s3})  &$62$ &  $45.3$    \\ \hline%

Triangular$(\bigtriangledown)$ (Fig.~\ref{s2}) &$55.6$ &  $42.2$    \\ \hline%

Embedded circle (~~\circle{6})(Fig.~\ref{chess6})   & $2.62$ & $1.86$   \\ \hline%

    \end{tabular}
    \caption{Transmission through checkerboard slab lens with embedded regions of different configurations
    for various values of material parameters.The last column shows the transmission that would have been obtained had a
    silver slab lens been used for imaging. The slab lens is seen to be the most efficient, while transmissions greater than 40 \%
    are obtained for the triangular inclusions, where no light is expected to be transmitted. The origami lens with circular inclusions
    shows a dramatic reduction in transmission, and transmits very little radiation, in close correspondence with its behaviour as predicted
    by ray diagrams, but in contradistinction with the generalized lens theorem. The transmittances slightly greater than 100 \%
    can be attributed to the lack of convergence of the calculations.}
   \label{tab1}}
   \vspace{0cm}\mbox{}
\end{table}



\begin{figure}[tbp]
\hspace{2cm}\mbox{}
\scalebox{0.7}{\includegraphics[angle = -0, width = 1.0\columnwidth]{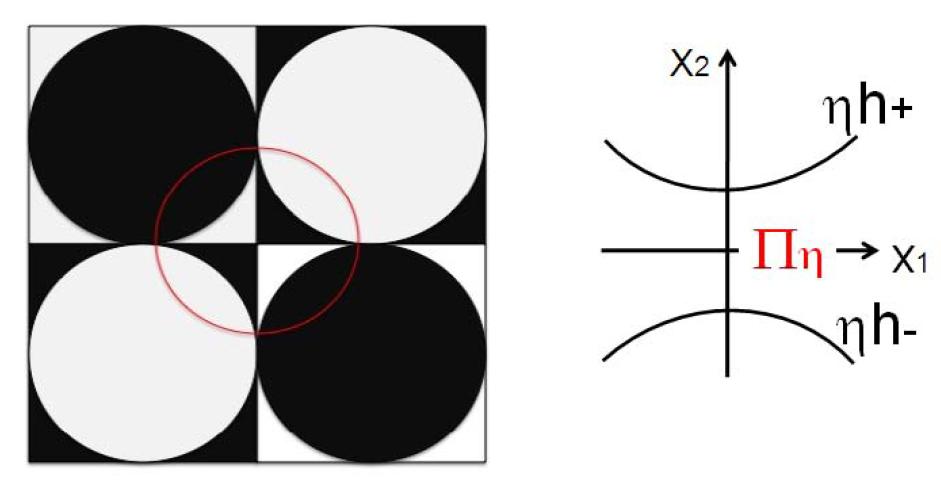}}
\caption{\label{circles}
Schematic diagram of the checkerboard with embedded circles, with the interstitial region highlighted in red
(left). The interstitial  space has been modelled as a large curved diamond-shaped region connected to four thin domains: These are curved ligaments bounded by two functions $h_{\pm}$ scaled by a small parameter $\eta$ (right).}
\label{thineta}
\end{figure}

These ones provide a reflectionless interface between the region of interest (a large middle square
containing the line source and the silver checkerboard on Fig.~\ref{s1}) and the PML (four elongated rectangles and
four small squares) at all incident angles. It is obvious from Figs.~ \ref{s1}, \ref{s2} that both rectangular and triangular checkerboards enable some imaging process in full contradiction with the ray picture. Extraordinary transmission is at work! There is a large concentration
of fields along the interfaces between positive and negative media, which depend upon the symmetry of the systems under consideration. This is exemplified by Figs.~\ref{s3} and ~\ref{s4}. Dissipation is seen to affect these checkerboards, particularly the ones with embedded
triangular and circular inclusions. The results obtained for the transmission characteristics of such checkerboards, both the non-dissipative case (with $\epsilon = -1, \mu = -1$) as well as dissipative ones with $\epsilon = -1 + i 0.01, \mu = 1$ and $\epsilon = -1 + i 0.1, \mu = 1$, have been summarized in
Table~\ref{tab1}. It is illuminating to interpret the data in this table using either the ray optics or wave viewpoints: the former tells us that 50 \% of
rays should be transmitted through the rectangular checkerboard lens, whereas all rays should be diffracted for the triangular checkerboard lens; On the contrary, the generalized lens theorem states that any such checkerboard lens has a full transmission in the limit of no dissipation. This wave picture
is indeed fully adequate for the ideal case whereby $\varepsilon=\mu=-1$ in the NRIM, as exemplified by the first column of
Table~\ref{tab1}. However, the situation is slightly different when we introduce some dissipation in the NRIM: The transmission hardly exceeds 50 \% for rectangular and triangular checkerboard silver lenses, as reported in the last column of Table~\ref{tab1}. Interestingly, the case of a checkerboard lens with embedded circles as in Fig.~\ref{chess6} is clearly beyond the scope of the generalized lens theorem: even a small dissipation leads to a dramatic drop in the transmittance, with less than 2 \% of light passing through such a silver lens (in contradistinction with the generalized lens theorem, but in agreement with the ray picture). Such a counter-example for the application of the perfect lens theorem to dissipative NRIM should warn us that the behaviour of structured perfect lenses with thin-bridges (within which oscillations of the electromagnetic field are tremendously enhanced) cannot be fully predicted within the frame of the theorem (and moreover numerical simulations should take into account the strongly enhanced non-linear effects). The reason for that is very simple: The interstitial space between the disks can be modelled as a large curved diamond (see Fig.~\ref{circles}) connected to four thin domains $\Pi_\eta=\{{\bf x}\in {IR^2}: \; l/2<x_1<l/2 \; , \; \eta h_{-}(x_1)<x_2<\eta h_{+}(x_1)\}$
see Fig. \ref{thineta}. Assuming that either the electric or magnetic field is orthogonal to the plane, the Maxwell equations reduce to
\begin{equation}
\Delta u + \omega^2/c^2 u = 0 \; ,
\end{equation}
in each homogenous region (shown in black and white in Fig.) of the Bridge, and the
the transmission conditions at the interface between these regions as well as on the upper and lower boundaries $h_{-}$ and $h_{+}$ of $\Pi_\eta$ is
given by the continuity of $u$ and a negative jump of its normal derivative i.e. $[u]=0$ and $[du/dn]=-1$. Assuming the following ansatz for $u$:
\begin{equation}
u\sim u_0(x_1) + \eta^2 u_1(x_1,\xi) \;
\end{equation}
where $\xi=x_2/\eta$ and the rescaled gradient
\begin{equation}
\nabla=\nabla_{x_1} + \frac{1}{\eta} \nabla_\xi \; ,
\end{equation}
the Helmholtz equation is reduced to an ordinary differential equation for $-l/2<x_1<l/2$ at the leading order:
\begin{equation}
\frac{d}{dx_1} \left((h_{+}(x_1)+h_{-}(x_1)) \frac{d}{dx} u_0\right) + \omega^2/c^2 (h_{+}(x_1)+h_{-}(x_1)) u_0 = 0
\; .
\end{equation}
This equation is supplied with boundary conditions at the endpoints $x_1=\pm l/2$. Such boundary
conditions can be easily derived from the divergence theorem applied to the flux through the interface between
each thin bridge $\Pi_\eta$ and the large region $\Sigma$ to which they are connected (keeping in mind the sign change across each interface).
In the case of constant curvature $a$, that is when $h_{-}=h_{+}=1+a^2h^2/2$ (here, thin-bridges are indeed interstitial spaces between circular regions), this leads the following resonant frequency (see eqn. (4.12) in ~\cite{prsa2007}:
\begin{equation}
\omega \sim 2\eta^2 \frac{c^2}{\rm{area}(\Sigma)} \frac{a h}{\hbox{atan}(a l/2)} \; .
\end{equation}

Compared to the earlier work in ~\cite{prsa2007}, such resonant frequencies are highly degenerate in the present case: in the limit
of zero dissipation in NRIM, $\hbox{area}(\Sigma)$ vanishes and the resonant frequencies $\omega$ form a continuum. This leads to a highly resonant structure at any frequency, which is a hallmark of an infinite local density of states.

\begin{figure}[tb]
\begin{center}
\scalebox{1.0}{\includegraphics[width=10cm,angle=-0]{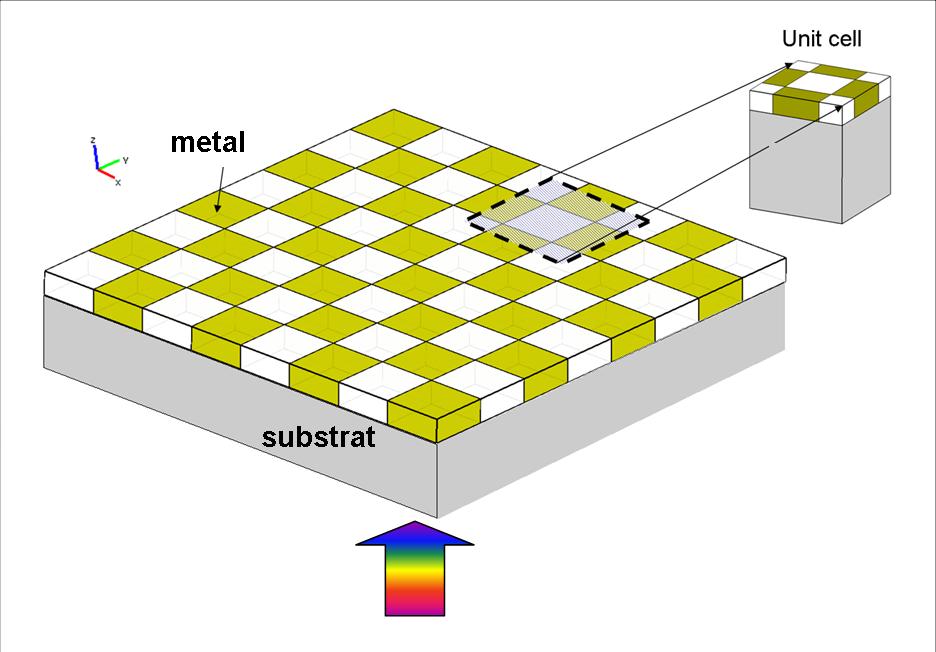}}
\caption{Diagrammatic view of the experimental setup: Left: Some white light
(shown with rainbow color in the large arrow)  illuminates a gold/silver checkerboard, and
transmission measurements are made. Right: Unit cell used in the numerical model.}
\label{muamer1}
\end{center}
\end{figure}

We now turn to the transmission properties of three-dimensional checkerboard structures, see Fig.~\ref{muamer1}. The generalized perfect lens theorem is still applicable to this configuration, which should then exhibit full transmittance in the limit of zero dissipation.
We have modelled such a transversely periodic structure using a three-dimensional unit cell, as depicted in the right panel of Fig.~\ref{muamer1}, with periodicity conditions on the vertical walls, and perfectly matched layers in the top and bottom layers.

\section{Experimental fabrication and measurements}

We report here on the experimental fabrication of photonic
checkerboard structures on gold films. Uniform gold films 200 nm
thick were deposited on polished fused silica substrates using dc
sputtering. The resulting films had a surface roughness of about
$\pm 2$ nm measured by atomic force microscopy.A dual beam focused ion 
beam (FIB) system (FEI NOVA 600)  equipped with a field emission a ion source
was used to create all the photonic structures presented here. 
Ion energies of about 30 keV and a beam current of about 10 pA were 
used for the nanopatterning. 

\begin{figure}
\hspace{2cm}\mbox{}
\scalebox{1.0}{\includegraphics[width=1.0\columnwidth,angle=-0]{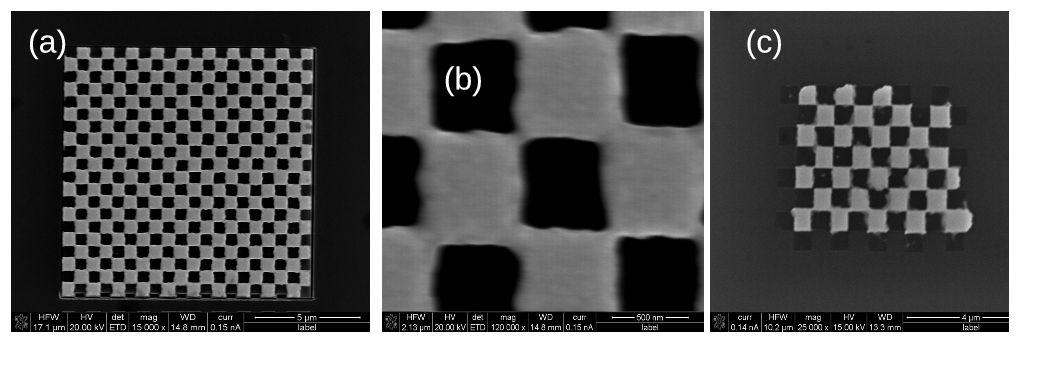}}
\vspace{-2cm}\mbox{}
\caption{Scanning electron microsope images of checkerboard structures made
by FIB induced adhesion
followed by peel-off on (a) a 200 nm thick gold film: each square has a size of about 
500nm $\times$ 500nm, (b) a magnified
picture of a part of the structure in (a) showing the connected nature of
the checkerboard” and (c) a 200 nm thick silver films with 
each checkerboard square measuring about 450nm $\times$ 450 nm.  \label{nanochess}}
\end{figure}

The checkerboards structures were fabricated by direct milling the gold films by the FIB. 
In order to retain only the checkerboard structures while removing the remainder of the
unpatterned gold film, we utilize the technique of focussed ion beam induced adhesion that has been
recently developed~\cite{neeraj_nimb2009}. The adhesion of a film on a given substrate 
depends on the nature of the substrate-film interface and surface energies. 
Gold poorly adheres to silica and usually a thin layer of chromium or titanium 
is used for proper adhesion. There is no such extra layer in our samples.
Irradiation with an ion beam can, however, enhance the adhesion of such a poorly
adhering film that has been observed earlier for gallium ions with
energies of 10-30 keV. Essentially, the focussed ion beam is used to irradiate the patterned 
regions that need to be retained where strong adherence is
created between the structure and the substrate, then followed by peel-off using adhesive tape
that remove the non-irradiated poorly adhering regions of the films. 
Using the focused ion-beam induced adhesion followed by peel-off, 
square checkerboard structures were retained for both gold and silver 
films. The scanning electron microscope pictures of these 
obtained structures  are shown in Fig.~\ref{nanochess}. It is noted that the peel off 
technique works well down to checkerboard squares of about 
250nm (period of 500nm). However, it has not been possible to make 
much smaller structures as they get torn during the peel-of process. In fact, 
the peel off process appears to be more suitable for structuring
a softer material like gold rather than silver. Fig.~\ref{nanochess}(c)
shows a small silver checkerboard and should be compared with Fig.~\ref{nanochess}(a)
which shows a similar sized pattern in gold. The tear marks in the silver structure
can be clearly discerned. Fig.~\ref{nanochess}(b) shows a magnified 
view of the gold checkerboard array showing the connected nature of the
checkerboard array that is formed. 

\begin{figure}[tb]
\begin{center}
\scalebox{1.0}{\includegraphics[width=10cm,angle=-0]{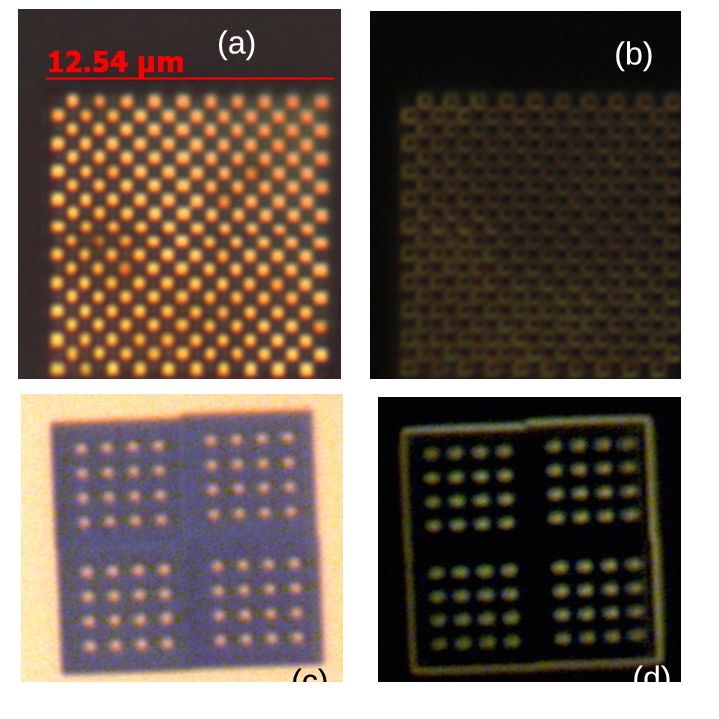}}
\caption{Optical (white light) microscope images of the gold checkerboard structures at 100X 
magnification. (a) shows the optical bright image image while (b) shows the dark-field image of the
checkerboard pattern shown in Fig.~\ref{nanochess}(a). Panel (c) shows the bright field image 
and (d) the dark field image of a sample with unconnected square scatterers with the same periodicity 
and the same scatterer size. In case of (d) each square scatterer appears uniformly bright in contrast
to (b). The film thickness is 200 nm in each case.}
\label{optical_nanochess}
\end{center}
\end{figure}

\begin{figure}[tb]
\begin{center}
\scalebox{1.0}{\includegraphics[width=10cm,angle=-0]{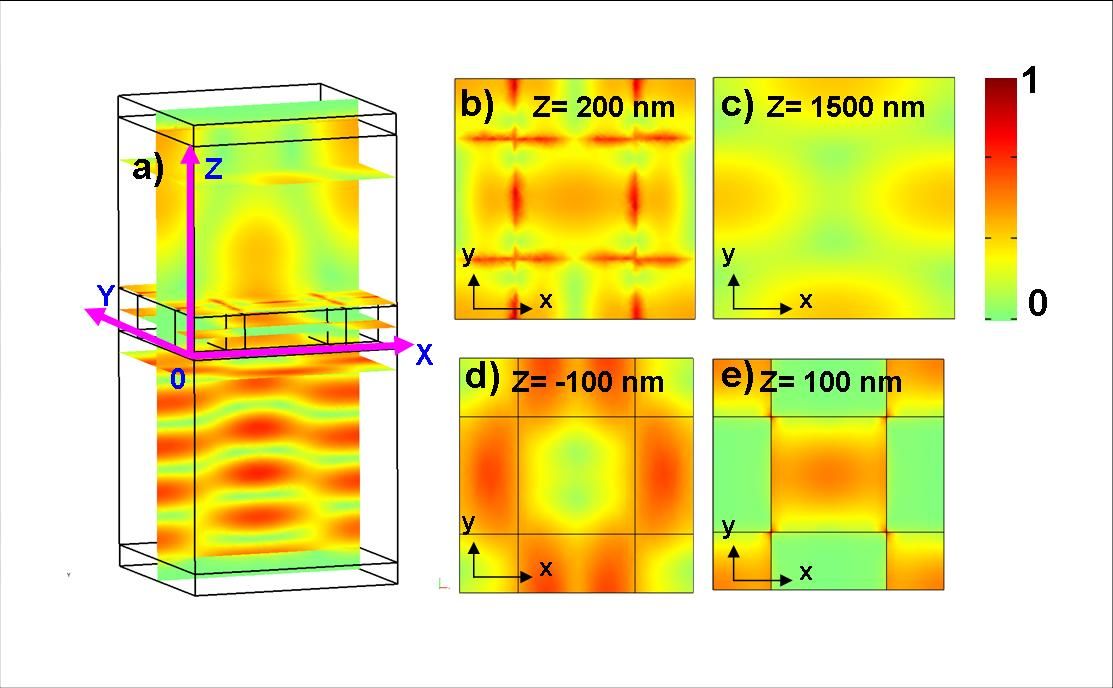}}
\caption{Three-dimensional COMSOL\texttrademark calculations for the propagation of light having
a wavelength of 580 nm across the unit cell shown in Fig.~\ref{muamer1}. We show the normalized total field of a 3d simulation for a periodic cell of sidelength 1000 nm and thickness 200 nm, with a square hole of sidelength 500 nm. The
field is incident through the substrate at 580nm (a). Three dimensional electric field plot (b), (c), (d) \& (e): cross section plots of the electric field corresponding to the different z-positions indicated on the figures.
}
\label{cboard_edge_fields_COMSOL}
\end{center}
\end{figure}

The scattering spectra of the checkerboard arrays were obtained by a
Ocean Optics USB 400 spectrometer through a optical fibre connected
to the trinocular port of a polarizing microscope (Olympus BX 51).
Fig.~\ref{optical_nanochess} shows the bright field optical images
of the checkerboard structures when viewed with a 100X objective.
For the case of the  $500\mathrm{nm} \times 500\mathrm{nm}$
checkerboards, each individual square is separately clearly visible
while for the $250\mathrm{nm} \times 250\mathrm{nm}$ checkerboards,
each square is not individually discernible because of the the
diffraction limit. Comparatively,  the dark field images in
Fig.~\ref{optical_nanochess} (b) interestingly show only the
outlines of the individual squares in the case of the
$500\mathrm{nm} \times 500\mathrm{nm}$ checkerboard array.
Essentially, the specular reflection has been cutoff in the
dark field images by using  light incident at very large angles compared
to the bright field images. In Fourier Optics
terms, the waves with $k\simeq 0$ have been removed by the dark
field filter. Thus, from the dark-field images we can conclude that
most of the scattering in the checkerboard system occurs at the
edges and corners of the system. This underlines our earlier
analyses emphasizing the importance of the corners and edges in the
system. For comparison, we show in
Fig.~\ref{cboard_edge_fields_COMSOL} the simulated fields excited by
a line source placed within the checkerboard which shows the
electric fields concentrated at the edges and corners of the
checkerboard structures. The simulations were performed using
the COMSOL$^\mathrm{TM}$ FEM solver. The bulk of the checkerboard
material within the squares hardly has any fields comparatively. In
order to cross-check this conclusion that the connected nature of
the checkerboard gives rise to the scattering from the edges and
corners, an array of unconnected square pads of gold and of the same
size ( $500\mathrm{nm} \times 500\mathrm{nm}$) was fabricated. In
this case, the entire square elements appear  uniformly bright in
both the bright field as well as the dark-field images [see
Fig.~\ref{optical_nanochess} (c) and (d)]. There is no evidence of
large scattering from the edges and corners in these unconnected structures.


\begin{figure}[tb]
\begin{center}
\hspace{1cm}\mbox{}
\scalebox{1.0}{\includegraphics[width=14cm,angle=-0]{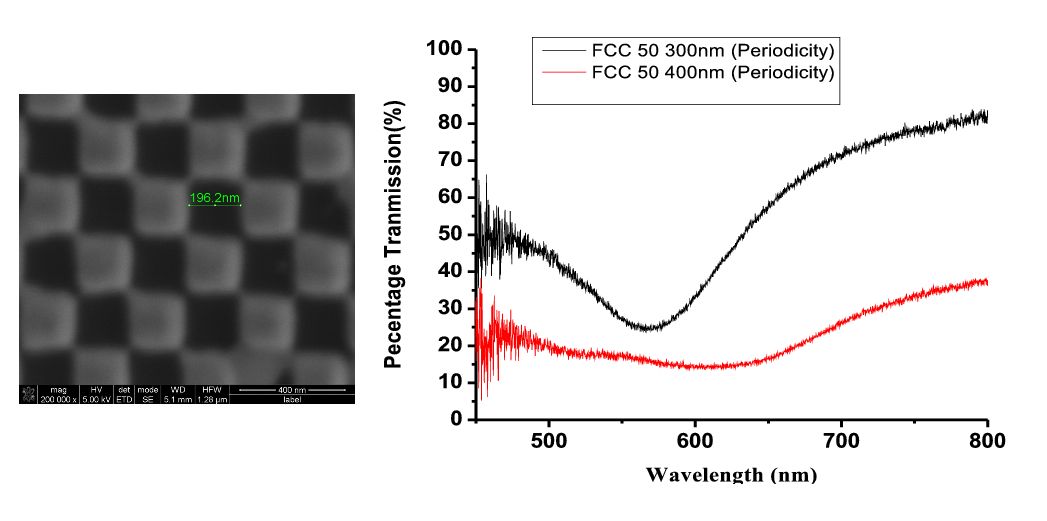}}
\vspace{-2cm}\mbox{}
\caption{Broadband extraordinary transmission 
through subwavelength square holes of size 
( $150\mathrm{nm} \times 150 \mathrm{nm}$ and  $200\mathrm{nm} \times 200 \mathrm{nm}$) 
placed in checkerboard fashion. The checkerboard has periodicity $300\mathrm{nm} \times 
300 \mathrm{nm}$ and   $400\mathrm{nm} \times 400 \mathrm{nm}$. There is broadband transmission
across the spectral region 600 nm to 900 nm. \label{eot_nanochess}}
\end{center}
\end{figure}

Finally, we measured the extra-ordinary transmission of light through subwavelength
square holes ( $150\mathrm{nm} \times 150 \mathrm{nm}$ and  
$200\mathrm{nm} \times 200 \mathrm{nm}$) placed in a checkerboard pattern 
(of periodicity $300\mathrm{nm} \times 300 \mathrm{nm}$ and
 $400\mathrm{nm} \times 400 \mathrm{nm}$)  that were fabricated by direct FIB milling
of a 200 nm thick film of gold (shown in Fig.~\ref{eot_nanochess}). Since the feature sizes are a fraction 
of the wavelength size at near-infra-red wavelengths, these checkerboards are expected to mimic
the action of checkerboards made of negative refractive index materials.
The transmission through a checkerboard structure of a given total area was normalized
by recording the direct transmission of light through a
hole of the same area as the patterned region in the otherwise optically opaque film.
These optically thick films with highly subwavelength
structures show a broadband extra-ordinary transmittance throughout the red to
the near infra-red spectrum (600 nm to 900 nm). Our apparatus does not allow us to
measure beyond 900 nm. We must remark here that such large
bandwidths are surprising and surmise here that the unique plasmonic properties of the
checkerboard structure leads to such properties. Essentially surface plasmon
excitations are possible at almost all frequencies below the plasma
frequency due to the very large degeneracy of the surface plasmon states
in the checkerboard systems. We believe that this degeneracy creates
the conditions for the broadband extraordinary transmission in this system.

\section{Conclusions}
We have presented here an assortment of the surprising photonic and
plasmonic properties of checkerboard structures of negative
refractive index materials and demonstrated some of the properties
in plasmonic checkerboard structures of gold. Checkerboards of media
with differing signs of material parameters result in  strange
landscapes for the plasmons that reside on the interfaces of the
system. There can be large concentration of electromagnetic fields
(local field enhancements), and large density of modes (large
scattering). But within the ambit of the generalized lens theorem,
the net sum effect on radiation of two equal sized but complementary
media is null. We have verified this result numerically for a
variety of complementary checkerboard systems. We have shown that
the numerical calculations on checkerboard systems are very
vulnerable to numerical artifacts and spurious resonances arise due
to finite differencing. Our experiments on plasmonic gold
checkerboard structures have shown that most of the scattering
appears to arise from the corners and edges of the system. We have
also demonstrated a surprisingly broadband extra-ordinary
transmission through subwavelength sized checkerboard structured
thick gold films.

While some authors consider a slab of NRIM as
Alice's mirror \cite{alice}, we may say NRIM checkerboards behave in
some way like the famous Alice's Cheshire cat who has the annoying
habit of disappearing and appearing at random times and places.
``Well, I've often seen a cat without a grin'', thought Alice, ``but a
grin without a cat is the most curious thing I ever saw in my
life''.

\section*{Acknowledgments}
The authors acknowledge funding from the Indo-French Centre for Promotion
of Advanced Research, New Delhi under grant no. 3804-02. The contribution
of the Inter University Centre, Indore where the gold films used for the study
were deposited is acknowledged.

\end{document}